\pdfoutput=1
\documentclass[journal=jctcce,manuscript=article,layout=twocolumn]{achemso}

\makeatletter
\let\l@addto@macro\relax
\makeatother
\usepackage[fontsize=10pt]{scrextend}
\usepackage{bm}
\usepackage[font=small,labelfont=bf]{caption}

\mciteErrorOnUnknownfalse

\usepackage{chemformula} 
\usepackage[T1]{fontenc} 
\usepackage{lipsum}
\usepackage[colorlinks,allcolors=blue]{hyperref}
\usepackage[capitalise]{cleveref}

\DeclareMathOperator*{\argmax}{arg\,max}
\DeclareMathOperator*{\argmin}{arg\,min}

\author{Muhammad R. Hasyim}
\affiliation{Simons Center for Computational Physical Chemistry, New York University, \\ New York, NY 10003, USA}

\author{Arkajit Mandal}
\affiliation{Department of Chemistry, Texas A\&M University, College Station, TX 77840, USA}
\author{David R. Reichman}
\affiliation{Department of Chemistry, Columbia University, New York, NY 10027, USA}
\email{mh7373@nyu.edu, mandal@tamu.edu, drr2103@columbia.edu}

\title[An \textsf{achemso} demo]
  {Towards Accurate Mixed Quantum Classical Simulations of Vibrational Polaritonic Chemistry}

\abbreviations{IR,NMR,UV}

\usepackage{xcolor} 

\newcommand{\dif}{\mathrm{d}}

\usepackage{titlesec}
\titleformat{\section}
  {\normalfont\fontsize{12}{15}\bfseries}{\thesection}{1em}{}

\allowdisplaybreaks

\begin{document}







\begin{abstract}
Interest in vibrational polaritonic chemistry, where ground-state chemical kinetics are modified via confined optical modes in a cavity, has surged in recent years. 
Although models have been developed to understand cavity-modified reactions, fully quantum mechanical simulations remain out of reach for the collective regime that involves many molecules, a critical aspect of the phenomenon. 
Mixed quantum-classical (MQC) simulations offer a scalable alternative, but their accuracy requires testing and potential improvements even in the single-molecule limit.
In this work, we take this step by first introducing the mapping approach to surface hopping (MASH) to address the limitations of traditional MQC methods. 
Second, we incorporate a quantum treatment of the cavity mode, moving beyond the classical approximations often employed in previous studies. 
Results for a single-molecule model of vibrational polaritonic chemistry show that combining MASH with a quantum cavity mode yields the most accurate rates. 
However, this scheme may produce different long-time population dynamics at zero coupling depending on whether the cavity mode is quantized; a problem known as size-inconsistency in MASH.
We address this problem proposing the $\epsilon$-MASH approach, which forbids hopping between states with negligible nonadiabatic couplings (NACs). 
Combining MASH with a quantum cavity mode thus provides a promising approach for scalable and accurate MQC simulations in the collective regime.
\end{abstract}

\section{Introduction} 
Polaritonic chemistry is a field that investigates the potential of new reactivities that arise when molecular systems are placed in structures that sustain confined optical modes, e.g., Fabry-Pérot cavities or plasmonic media.\citep{dunkelberger2022vibration}
Recent experiments have reported that the rate of ground-state chemical reactions can be either accelerated or slowed down by aligning the frequency of a cavity mode with a molecular vibrational frequency.\citep{thomas2016ground,thomas2019tilting,lather2019cavity,vergauwe2019modification,lather2020improving,hirai2020modulation,ahn2023modification,lather2022cavity}
At this resonance condition, the system enters a vibrational strong coupling (VSC) regime and forms hybrid light-matter states known as polaritons.
VSC is characterized by a distinct spectroscopic feature called the Rabi splitting, where the original frequency splits into two peaks: the upper and lower polariton frequencies. 
The size of this Rabi splitting, denoted as $\Omega_R$, reflects the coupling strength, with larger splittings indicating stronger coupling. 

An open issue in polaritonic chemistry is the development of a satisfactory theory for understanding how VSC modifies chemical reactions.\citep{mandal2023theoretical} 
The main theoretical challenge lies in the inherently collective nature of the phenomena: experiments suggest that a macroscopic number of molecular degrees of freedom when {\it collectively} coupled to the cavity radiation leads to modification of chemical reactivity. Although it is simple to show that collective light-matter coupling leads to a Rabi splitting $\Omega_R $ that scales as $\sqrt{N}$, as is observed in experiments~\citep{ThomasNp2020, mandal2023theoretical}, how  a cavity can alter chemical kinetics remains a mystery to
date. 
Accurate simulations of light-matter hybrid systems are crucial to understanding this phenomena. 
However, reliably modeling the dynamics of many molecules beyond the mean-field ($N \to \infty$) limit\citep{fowler2022efficient,fowler2024mean,lindoy2024investigating} in an accurate way is computationally intractable within a fully quantum framework, necessitating approximations.
Mixed quantum-classical (MQC) simulations\citep{crespo2018recent} offer a practical approach by treating molecular vibrations quantum mechanically while describing other degrees of freedom (DOFs) classically. 
Using this approach, recent work\citep{hu2023resonance} has shown that a classical cavity mode suffices to capture the qualitative trends of resonant reaction rate enhancements in the single-molecule limit. 
However, MQC methods tend to overestimate the enhancement by an order of magnitude or more.\citep{hu2023resonance}
This limited accuracy, even at the single-molecule limit, remains an obstacle for scalable and accurate simulations of vibrational polaritonic chemistry. 

In this work, we investigate strategies to improve the accuracy of MQC methods in a prototypical model\citep{hu2023resonance,lindoy2023quantum} for VSC-modified chemistry. \Cref{sec:model} introduces the model, describing a single molecule interacting with a lossy cavity mode and a solvent bath. 
We then present two approaches for enhancing MQC calculations of reaction rates. 
In \cref{sec:mash}, we introduce the mapping approach to surface hopping (MASH),\citep{mannouch2023mapping,runeson2023multi} an MQC approach that integrates the strengths of semiclassical mapping and surface hopping methods.
In \cref{sec:qph},  we redefine the system Hamiltonian using a quantum cavity mode to fully capture the reaction mechanism, including cavity-mediated enhancements, within a quantum mechanical framework. 
Results are presented in \cref{sec:results}, where we show that combining MASH with a quantum cavity mode leads to the most accurate reaction rate values. 

\begin{figure}[t]
    \centering
    \includegraphics[width=\linewidth]{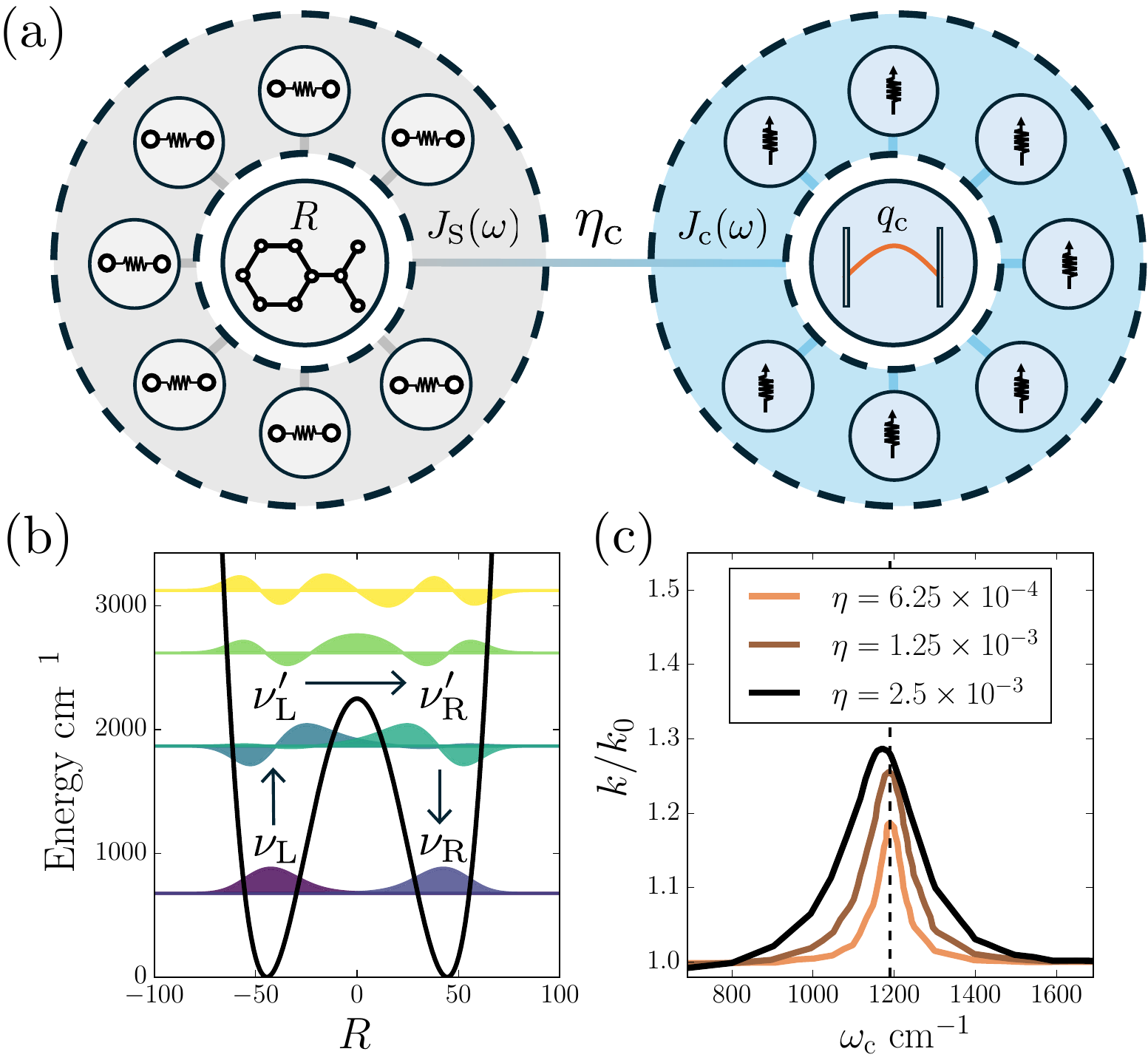}
    \caption{(a) Illustration of the model: A single molecule surrounded by a solvent bath is coupled to a lossy cavity mode. (b) Illustration of the double well along the reaction coordinate $R$ with the first six eigenstates shown. Arrows indicate the reaction mechanism outside of a cavity. (c) The reaction rate enhancement at the resonant frequency $\omega_0$, calculated with HEOM\citep{tanimura1990nonperturbative,tanimura2006stochastic,xu2005exact,xu2007dynamics} at different coupling strengths. Adapted from Ref.~\citenum{hu2023resonance}.}
    \label{fig:doublewell}
\end{figure}

\section{Model System} \label{sec:model}

In this work, we study a prototypical model for VSC-modified chemistry,\citep{hu2023resonance,lindoy2023quantum} working in atomic units where $\hbar = 1$; see \cref{fig:doublewell}(a) for an illustration. 
A single molecule with mass $M$ and reaction coordinate $\hat{R}$ is coupled to a cavity mode of frequency $\omega_\mathrm{c}$. 
The reaction coordinate Hamiltonian is $\hat{H}_\mathrm{R}= \frac{\hat{P}^2}{2 M}+ \hat{V}(\hat{R})$ with $\hat{P}$  the momentum operator and the potential energy $\hat{V}(\hat{R})$ of the form
\begin{equation}
\hat{V}(\hat{R}) = \frac{M^2 \omega_\mathrm{b}^4}{ 16 E_\mathrm{b}} \cdot \hat{R}^4   -\frac{1}{2} M \omega_\mathrm{b}^2 \cdot \hat{R}^2   \,,
\end{equation}
where $\omega_\mathrm{b}$ and $E_\mathrm{b}$ are the barrier frequency and barrier height, respectively.
Additionally, solvent molecules are modeled via a harmonic bath coupled to the reaction coordinate with the Hamiltonian $\hat{H}_\mathrm{S}$ written as 
\begin{equation}
   \hat{H}_\mathrm{S} = \sum_{i} \frac{\hat{p}_i^2}{2 m_i }+ \frac{1}{2} m_i \omega_i^2 \left(\hat{q}_i-\frac{g_i}{\omega_i^2} \hat{R} \right)^2 \,,
\end{equation}
where $\{ \hat{p}_i,\hat{r}_i\}$ are a set of momentum and position coordinates, and $\{ g_i, \omega_i \}$ are a set of coupling constants and bath frequencies. 
The coupling constants and frequencies are set by the spectral density $J_\mathrm{S}(\omega)$. 

The minimal coupling of the molecule to the cavity mode is described by the Pauli-Fierz (PF) Hamiltonian. The coupling strength is controlled by a parameter $\eta_\mathrm{c}$ and the PF Hamiltonian can be written within the dipole gauge and long wavelength approximation as
\begin{equation}
\hat{H}_\mathrm{PF} = \hat{H}_\mathrm{R} + \frac{\hat{p}_\mathrm{c}^2}{2}+\frac{\omega_\mathrm{c}}{2} \left(\hat{q}_\mathrm{c}+\sqrt{\frac{2}{\omega_\mathrm{c}}} \eta_\mathrm{c} \hat{R}\right)^2 \,, \label{eq:pf_hamiltonian}
\end{equation}
where $\hat{q}_\mathrm{c}$ and $\hat{p}_\mathrm{c}$ are position and momentum coordinates of the cavity mode, respectively. 
Lastly, the cavity mode couples to an additional harmonic bath whose Hamiltonian is 
\begin{equation}
\hat{H}_\mathrm{C} = \sum_{i} \frac{\hat{\pi}_i^2}{2 \tilde{m}_i}+ \frac{1}{2} \tilde{m}_i \tilde{\omega}_i^2 \left(\hat{r}_i-\frac{\tilde{g}_i}{\tilde{\omega}_i^2} \hat{q}_\mathrm{c} \right)^2 \,.
\end{equation}
Similar to the solvent, the cavity bath is described by a set of momentum and position coordinates $\{ \hat{\pi}_i,\hat{r}_i\}$ as well as a set of coupling constants and bath frequencies $\{ \tilde{g}_i, \tilde{\omega}_i \}$. 
The spectral density of both the solvent and cavity bath follow Drude-Lorenz forms, $J_\mathrm{S}(\omega)$ and $J_\mathrm{C}(\omega)$, with respective characteristic frequencies $\gamma_s$, $\gamma_c$ and reorganization energies $\lambda_s$, $\lambda_c$; see \cref{sec:modelparam} for parameter details.
In total, the full Hamiltonian is $\hat{H} = \hat{H}_\mathrm{PF}+\hat{H}_\mathrm{C}+\hat{H}_\mathrm{S}$. 

Outside of the cavity, the reaction can be understood from the diabatized version of its vibrational eigenstates $| \nu_k \rangle$, which satisfy $E_k | \nu_k \rangle = \hat{H}_\mathrm{R} | \nu_k \rangle$.\citep{hu2023resonance,lindoy2023quantum} Diabatizing the lowest two eigenstates yields $| \nu_L \rangle = \frac{1}{\sqrt{2}} \left(|\nu_0\rangle + |\nu_1\rangle \right)$ and $|\nu_R \rangle = \frac{1}{\sqrt{2}} \left(|\nu_0\rangle - |\nu_1\rangle\right)$, which are degenerate and localized to the left and right well, while the next two excited states transform into $| \nu_L^\prime \rangle = \frac{1}{\sqrt{2}} \left(|\nu_2\rangle + |\nu_3\rangle \right)$ and $|\nu_R^\prime \rangle = \frac{1}{\sqrt{2}} \left(|\nu_2\rangle - |\nu_3\rangle\right)$, which are degenerate but delocalized; see \cref{fig:doublewell}(b) for an illustration. The reaction mechanism proceeds from $| \nu_L \rangle$ through excitation to $| \nu_L^\prime \rangle$, followed by tunneling to $| \nu_R^\prime \rangle$ and relaxation to $| \nu_R \rangle$.
Inside a cavity, the rate limiting step ($| \nu_L \rangle  \to | \nu_L^\prime \rangle$) can be facilitated by the cavity mode acting as a rate-promoting vibrational mode, resulting in rate enhancement when $\omega_\mathrm{c}$ matches the $|\nu_L \rangle \to |\nu_L^\prime \rangle$ transition frequency $\omega_0$. 
In \cref{fig:doublewell}(c), we present the benchmark quantum-mechanical calculations based on hierarchical equations of motion (HEOM)\citep{tanimura1990nonperturbative,tanimura2006stochastic,xu2005exact,xu2007dynamics} that shows modest enhancements for a range of coupling strengths $\eta_\mathrm{c}$.


With the fundamental physics established, we introduce the first strategy for rate calculations, the mapping approach to surface hopping (MASH),\citep{mannouch2023mapping,runeson2023multi,lawrence2024size} by first introducing the general idea behind MQC methods.  

\section{MQC Dynamics with MASH} \label{sec:mash}

For a given nonadiabatic problem, we can assign classical variables in a quantum system through various approaches, typically resulting in a Hamiltonian $\hat{H}(\mathbf{q},\mathbf{p}) = \sum_{j} \frac{p_j^2}{2 m_j} + \hat{E}(\mathbf{q})$, where $\mathbf{q}$ and $\mathbf{p}$ are vectors of classical coordinates and momenta with elements $q_j$ and $p_j$, and $\hat{E}(\mathbf{q})$ contains both the classical potential energy $V(\mathbf{q})$ as well as the kinetic $\hat{T}$ and potential $\hat{V}(\mathbf{q})$ energy operators. In the diabatic basis where a quantum state $| \psi \rangle = \sum_{\mu} c_\mu | \mu \rangle$, the energy operator matrix elements are $E_{\mu \nu}(\mathbf{q}) = V(\mathbf{q}) \delta_{\mu\nu} + T_{\mu \nu}+V_{\mu \nu}(\mathbf{q})$ and the quantum subsystem evolves through a time-dependent Schrodinger equation (TDSE)
\begin{equation}
\dot{c}_\mu =-\mathrm{i} \sum_{\nu} E_{\mu \nu}(\mathbf{q}) c_\nu \,,
\end{equation}
while the classical system follows Newton's equations of motion
\begin{equation}
\dot{q_j} = p_j/m_j \,, \quad \dot{p_j} = F_j(\mathbf{q}) \,,
\end{equation}
where the specific mapping of $E_{\mu \nu}(\mathbf{q})$ to the force $F_j(\mathbf{q})$ distinguishes different MQC schemes. 
Various schemes are also distinguished by how they initialize the quantum and classical subsystems, though the latter often employs the thermal Wigner distribution to account for nuclear quantum statistics.

An important class of MQC methods used in polaritonic chemistry is semiclassical mapping approaches where the force $F_j$ is computed as an average the current quantum state. For instance, Ehrenfest dynamics expresses this force as
\begin{equation}
F_j = -\langle \psi | \nabla_{j} \hat{E}(\mathbf{q}) |\psi \rangle \,. \label{eq:mfeforce}
\end{equation}
Other approaches like $\gamma$-SQC\citep{cotton2019trajectory} yield similar force expression as \cref{eq:mfeforce} by mapping the quantum Hamiltonian onto the classical Meyer–Miller–Stock–Thoss (MMST) Hamiltonian\citep{meyera1979classical,stock1997semiclassical,thoss1999mapping,miller2016classical}.
The classical dynamics of the MMST Hamiltonian can produce unphysical outcomes, such as negative quantum state populations, which $\gamma$-SQC addresses through a windowing scheme.\citep{cotton2019trajectory} 
On the other hand, approaches based on spin mapping\citep{runeson2019spin,runeson2020generalized} represent two-level quantum systems as a vector on the Bloch sphere---a scheme that also generalizes to multi-state systems\citep{runeson2020generalized}---and uses stochastic sampling to initialize the quantum system, even if the initial condition is a pure state. 
Note that these methods have been applied to the current model,\citep{hu2023resonance} but their rate enhancement predictions deviate by a factor of four to eight, highlighting the need for alternative methods.

Another prominent class of MQC methods is surface hopping,\citep{tully1990molecular} where the classical force arises from the potential energy of the so-called active quantum state rather than a dynamical average. 
Formulated on an adiabatic basis, where $\hat{E}(\mathbf{q})| m; \mathbf{q} \rangle = E_m(\mathbf{q}) | m; \mathbf{q} \rangle$, the classical force due to an active state $|a; \mathbf{q}\rangle$ is given by
\begin{equation}
F_j = -\langle a; \mathbf{q} | \nabla_j \hat{E} (\mathbf{q}) | a; \mathbf{q} \rangle = -\nabla_j E_a(\mathbf{q}) \,, \label{eq:shforce}
\end{equation}
and the quantum state $| \psi \rangle = \sum_m c_m | m; \mathbf{q} \rangle$ evolves via the TDSE in the adiabatic basis
\begin{equation}
\dot{c}_m=-\mathrm{i} E_m(\mathbf{q}) c_m-\sum_j \frac{p_j}{m_j} \sum_n d_{m n}^j(\mathbf{q}) c_n \,, \label{eq:tdse_ad}
\end{equation}
where $d_{mn}^j(\mathbf{q})$ is the $j$-th component of the non-adiabatic coupling vector between state $m$ and $n$, with second-order coupling terms neglected.
Surface hopping uses stochastic state switching, guided by transition probabilities $P_{m \to n}$, to explore different potential energy surfaces. 
These probabilities define schemes like fewest-switches surface hopping (FSSH)\citep{tully1990molecular} and global flux surface hopping (GFSH)\citep{wang2014global}.
In the current model, surface hopping, particularly GFSH, predicts a rate enhancement that deviates more than $\gamma$-SQC and spin mapping approaches.\citep{hu2023resonance} 

Since neither mapping nor surface hopping approaches suffice for rate calculations in the current model, we adopt the mapping approach to surface hopping (MASH),\citep{mannouch2023mapping,runeson2023multi} which combines features of both methods. 
Like surface hopping, the classical force is derived from a single active quantum state, deterministically selected as the state with the highest instantaneous population. 
MASH also incorporates key elements of mapping approaches, such as sampling of initial wavefunction coefficients and formulas for estimating key observables, e.g., average populations and coherences. 

Initially formulated for two-state systems,\citep{mannouch2023mapping} MASH extends to multi-state systems in a couple of ways.\citep{runeson2023multi,lawrence2024size} 
Here, we use the multi-state MASH developed in Ref.~\citenum{runeson2023multi} that defines the active state $a$ as the adiabatic state $m$ with the highest instantaneous population, $\rho_m = c_m^* c_m$, i.e., $a = \argmax_{m} \{ \rho_m \}$.
The quantum system evolves via \cref{eq:tdse_ad} while the classical force in MASH is 
\begin{gather}
F_j = -\nabla_j E_a(\mathbf{q}) - \sum_{n} \Delta E_{na}(\mathbf{q})\delta(\mathbf{q} - \mathbf{q}_{na}) \delta_{na}^j \,,
\end{gather}
where $\Delta E_{na} (\mathbf{q})=E_n(\mathbf{q})-E_a(\mathbf{q})$, $\delta_{an}^j = \nabla_j (\rho_n - \rho_a)$, and $\mathbf{q}_{na}$ is the system configuration when $\rho_n=\rho_a$. Note that the first term, $-\nabla_j E_a(\mathbf{q})$, corresponds to the classical force in surface hopping (\cref{eq:shforce}).
The second term represents an impulse force that switches on when the system hops from the current state $a$ to a new state $n$, with $\mathbf{q}_{an}$ the system configuration at the hopping threshold.
Physically, the impulse corresponds to the classical system encountering a step potential barrier of height $\Delta E_{an}$ when $\rho_n=\rho_a$. 
If the system's kinetic energy exceeds this barrier, it resizes its momentum projected on $\delta^j_{an}$ to conserve the total energy and switches the active state to $n$. 
Otherwise, the momentum reverses, and the system remains in state $a$.

MASH has succeeded in spin-boson systems,\citep{mannouch2023mapping} the Marcus electron transfer problem,\citep{lawrence2024recovering} and the FMO model.\citep{mannouch2023mapping,runeson2023multi} 
The similarities between these previously tested systems and the current polaritonic problem suggest that MASH can achieve similar results. The next step is to determine which parts of the system are treated classically vs. quantum mechanically.



\section{Quantum Cavity Mode} \label{sec:qph}

To define the quantum subsystem, we first consider the diabatic states constructed from the vibrational eigenstates.
Due to the nature of the reaction dynamics, we can work in a truncated subspace spanned by $\{| \nu_L \rangle, | \nu_R \rangle, |\nu_L ^\prime\rangle, | \nu_R^\prime \rangle\}$, as defined in \cref{sec:model}. 
We relabel this basis set as $\{ | \mu \rangle \}_{\mu=1}^4$ for convenience and express the reaction coordinate Hamiltonian as $\hat{H}_\mathrm{R} = \sum_{\mu,\nu} (\mathbf{H}_\mathrm{R})_{\mu \nu} |\mu \rangle \langle \nu |$; see \cref{sec:matelems} for explicit elements of the matrix $\mathbf{H}_\mathrm{R}$. From this point, we have two possible routes. The first, following previous work,\citep{hu2023resonance} treats the cavity mode, loss, and solvent bath as classical DOFs. While this approach captures the resonant enhancement effect, its accuracy is limited.\citep{hu2023resonance}
The second approach we pursue here quantizes the cavity mode using Fock states. 
To introduce Fock states, we introduce creation ($\hat{a}^\dagger$) and annihilation ($\hat{a}$) operators, which express the cavity-mode position and momentum variables as $\hat{q}_c=\sqrt{\frac{1}{2 \omega_c}}\left(\hat{a}+\hat{a}^{\dagger}\right)$ and $\hat{p}_{\mathrm{c}} = \mathrm{i} \sqrt{\frac{\omega_c}{2}}\left(\hat{a}^{\dagger}-\hat{a}\right)$. 
Substituting them into the PF Hamiltonian in \cref{eq:pf_hamiltonian}, we obtain 
\begin{gather}
\hat{H}_\mathrm{PF} = \hat{H}_\mathrm{R} + \hat{H}_\mathrm{p} +\hat{H}_\mathrm{int} \,, \label{eq:pf_h_aadagger}
\\
\hat{H}_\mathrm{p} = \omega_\mathrm{c} \left(\hat{a}^\dagger \hat{a}+\frac{1}{2} \right) \,,  \label{eq:freephoton_h}
\\
\hat{H}_\mathrm{int} = \omega_\mathrm{c} \eta_\mathrm{c} \hat{R} (\hat{a}+\hat{a}^\dagger) + \omega_\mathrm{c}\eta_\mathrm{c}^2 \hat{R}^2 \,,
\label{eq:pf_hint}
\end{gather}
where $\hat{H}_\mathrm{p}$ is the free-photon Hamiltonian and $\hat{H}_\mathrm{int}$ is the interaction term. The Fock states of the cavity mode are then defined by the diagonalization $\left(N +\frac{1}{2}\right)| N \rangle = \left(\hat{a}^\dagger \hat{a}+\frac{1}{2} \right)   | N \rangle$, with $N$ representing the number of photons. 
With a quantum cavity mode, the quantum subsystem now becomes a Kronecker product between the four vibrational eigenstates and $N_\mathrm{p}$ photonic Fock states. 
See \cref{sec:matelems} for the explicit matrix elements of \cref{eq:pf_hint,eq:freephoton_h,eq:pf_hint}. 

\begin{figure}[t]
    \centering
    \includegraphics[width=\linewidth]{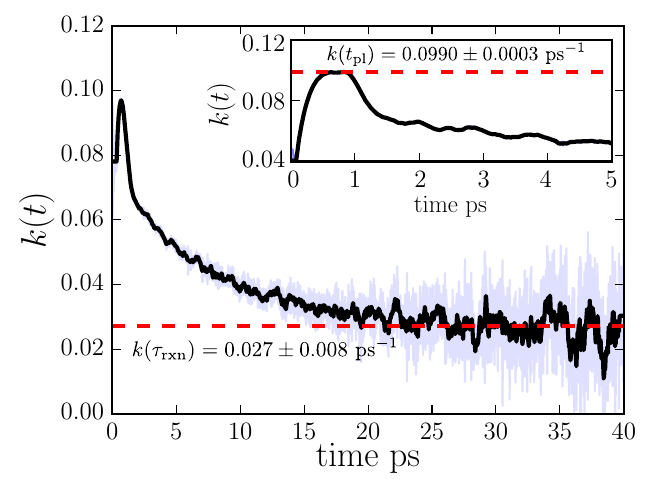}
    \caption{Plot of $k(t)$ using Ehrenfest dynamics off-resonance  where at short times (inset) a flat peak emerges representing a brief plateau region at $t=t_\mathrm{pl} \sim 1 \ \mathrm{ps}$, and at long times a new plateau emerges at $t \gg \tau_\mathrm{rxn}$. Shaded blue region is 99\% confidence interval.}
    \label{fig:howtocalc}
\end{figure}

\begin{figure*}[t]
    \centering
    \includegraphics[width=\linewidth]{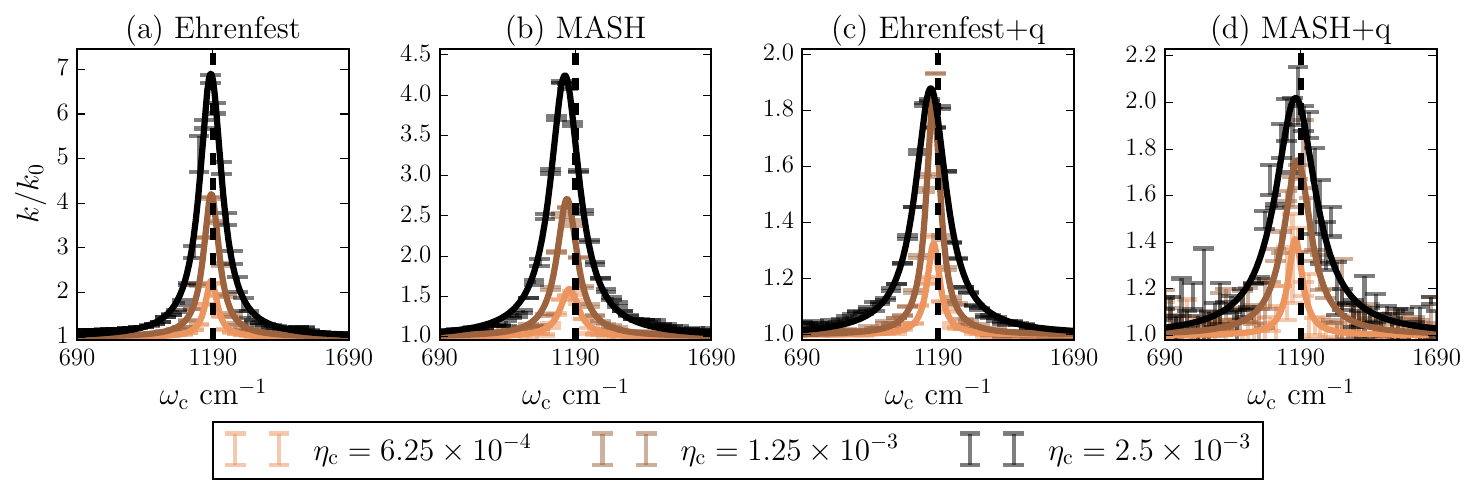}
    \caption{The rate profile as a function of cavity-mode frequency $\omega_\mathrm{c}$ calculated with Ehrenfest (a) and MASH (b) with a classical cavity mode. The corresponding rate profiles using a quantum cavity mode are found in (c) and (d), respectively. The dashed vertical line corresponds to the resonant frequency at $\omega_0 = 1190 \ \mathrm{cm}^{-1}$. Error bars represent 99\% confidence intervals.}
    \label{fig:ratesummary}
\end{figure*}

\begin{figure}
    \centering
    \includegraphics[width=\linewidth]{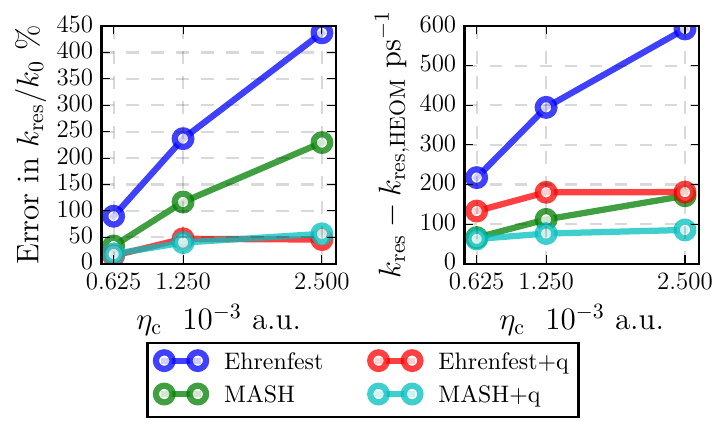}
    \caption{(Left) The percent error in the predicted rate enhancement and (right) the difference in rate values at resonance, compared to HEOM.}
    \label{fig:error}
\end{figure}

Convergence to the $N_\mathrm{p} \to \infty $ limit becomes increasingly challenging at strong coupling strengths $\eta_\mathrm{c} $ due to cavity mode polarization. 
Here, polarization refers to the displacement of the cavity mode from its $|0\rangle $ ground state, driven by the permanent dipole moments that persist at the reactant and product wells. 
In the model, the dipole moment operator $\hat{d} \sim \hat{R} $ thus we see the cavity-mode position $\hat{q}_\mathrm{c} $ is displaced by $-\hat{q}_0 = \sqrt{\frac{2}{\omega_\mathrm{c}}} \eta_\mathrm{c} \hat{R} \sim \eta_\mathrm{c} \hat{d}$.
At large $\eta_\mathrm{c}$, polarization implies the inclusion of more Fock states, and prior work\citep{mandal2020polarized} has shown convergence requiring $N_\mathrm{p} > O(10^2)$, making calculations potentially expensive in our model.
To allow for more rapid convergence, we follow Ref.~\citenum{mandal2020polarized} and apply a two-step transformation on the system:
\begin{enumerate}
    \item The diabatic basis is first transformed into the Mulliken-Hush (MH) basis. By diagonalizing the reaction coordinate operator $\hat{R} |\mu^\prime\rangle = R_{\mu^\prime} |\mu^\prime\rangle$ we can write the transformation to the MH states as $ |\mu^\prime\rangle = \hat{U}_\mathrm{MH} |\mu\rangle $. 
    
    \item We then use the polaron transform defined as $ \hat{U}_\mathrm{PL} = e^{\mathrm{i} \hat{q}_0 \hat{p}_\mathrm{c}} $ and apply it to the full Hamiltonian, yielding $ \hat{H}^\prime = \hat{U}_\mathrm{PL}^\dagger \hat{H} \hat{U}_\mathrm{PL} $. Physically, the polaron transform dresses photons with the polarization, converting Fock states into polarized Fock states.\citep{mandal2020polarized}. 
    In practice, the polaron transform removes the displacement $ \hat{q}_0 $ from $\hat{H}$ and introduces overlap integrals in the off-diagonal matrix elements; see \cref{sec:matelems} for their expressions. 
\end{enumerate}
As shown in \cref{fig:loweta_eig,fig:mideta_eig,fig:higheta_eig} of \cref{sec:matelems}, this transformation enhances eigenspectrum convergence, enabling use of a single-excitation subspace at strong coupling, underscoring its utility for the model.

\section{Results and Discussions} \label{sec:results}

Before testing the proposed strategies, we review the basics of calculating reaction rates from MQC simulations. 
First, recall that reaction rate theory\citep{chandler1978statistical,chandler1987introduction} posits first-order kinetics for the reactant $P_R$ and product $P_P$ populations so that 
\begin{equation}
\frac{\dif  P_R}{\dif t} = k_{P \to R} P_P - k_{R \to P} P_R \,, \label{eq:1strate}
\end{equation}
where $k_{P \to R}$ and $k_{R \to P}$ is the backward and forward rate, respectively.
Under the detailed balance condition ($k_{P \to R} P_P^\mathrm{eq} = k_{R \to P} P_R^\mathrm{eq}$, where $P_P^\mathrm{eq}$ and $P_R^\mathrm{eq}$ are equilibrium product and reactant populations), $P_R$ decays exponentially to $P_R^\mathrm{eq}$ with a reaction time constant $\tau_\mathrm{rxn} = \frac{1}{k_{P \to R}}+\frac{1}{k_{R \to P}}$.
Using the reactive flux formalism,\citep{chandler1978statistical} we can relate the proposed kinetics to equilibrium autocorrelation functions, providing a time-dependent rate estimator $k(t)$ that can be calculated from molecular simulations.
If the rate law holds, $k(t)$ plateaus at $t_\mathrm{p}$, where $\tau_\mathrm{mol} \ll t_\mathrm{p} \ll \tau_\mathrm{rxn}$; here, $\tau_\mathrm{mol}$ is the molecular timescale for product-state fluctuations. 
If the separation of timescales does not hold, then the dynamics do not rigorously obey the posited reaction kinetics. 


\begin{figure*}[t]
    \centering
    \includegraphics[width=\linewidth]{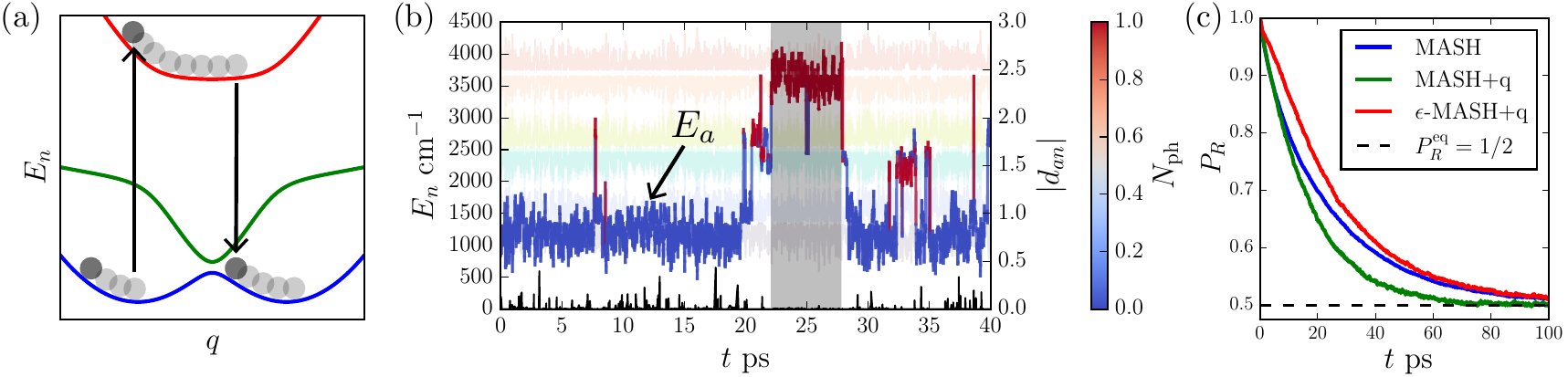}
    \caption{(a) Multi-state MASH\citep{runeson2023multi} can enable hopping between uncoupled adiabats, illustrated here with the first and third adiabats, ignoring the strong coupling region between the first and second adiabats.\citep{runeson2023multi} (b) Time series of adiabatic energies in MASH+q at zero coupling strength ($\eta_\mathrm{c}=0.0$), with the active-state energy $E_\mathrm{a}$ color-coded by photon number. Black lines correspond the the scalar nonadiabatic coupling $d_{an}$. Shaded region in grey shows a hop generating a single-photon excited state despite vanishing $d_{an}$, highlighting unphysical excitations facilitated by multi-state MASH. (c) Population dynamics for MASH at $\eta_\mathrm{c}=0.0$ with classical cavity mode, quantum cavity mode, and thresholded ($\epsilon$-MASH+q) hopping.}
    \label{fig:mashqproblem}
\end{figure*}

MQC methods present new complications as most fail to obey detailed balance or time reversibility, preventing the use of the reactive flux formalism. 
Instead, we rely on nonequilibrium simulations of population dynamics initiated from the reactant state. 
Using the Heaviside operator $\hat{h}_R(t)$ to project states to the reactant well, the reactant population is expressed as $P_R(t) = \mathrm{Tr}[\hat{h}_R \hat{\rho}(t)]$, where $\hat{\rho}(t)$ is the density matrix of the quantum subsystem. 
A time-dependent rate estimator $k(t)$ can be derived from \cref{eq:1strate} as:
\begin{equation} 
k(t) = \frac{\dot{P}_R(t)}{1 - P_R(t)/P_R^\mathrm{eq}} \,. 
\end{equation}
Note that errors from approximating the quantum dynamics affect how the plateau region emerges in $k(t)$. 
At short times, a plateau emerges as the system first commits to the reactant state, while at long times, a second plateau appears due to accumulated errors in the dynamics, as exemplified in \cref{fig:howtocalc}(a) by Ehrenfest dynamics.\protect\footnote{In this figure, the plateau region obtained from Ehrenfest lasts for 1 ps, which is brief. This result is not true for all methods, as our MASH calculations show a plateau of nearly 10 ps.}
Following previous studies,\citep{xie2013calculation,lawrence2024recovering} we calculate rates from the short-time plateau of $k(t)$, a regime where MQC methods more accurately approximate the quantum dynamics. 
This approach aligns with previous studies showing that various MQC methods like Ehrenfest, FSSH, and MASH can only reproduce analytical results for benchmark problems, e.g., Marcus electron transfer, when using the short-time plateau.\citep{xie2013calculation,lawrence2024recovering}



With the rate calculation procedure established, \cref{fig:ratesummary} summarizes the results using the introduced strategies, fitted to an analytical curve based on the Fermi's golden rule.\citep{ying2023resonance} Across all methods, the predicted rate enhancement grows more sharply with increasing coupling strength compared to the benchmark HEOM prediction (\cref{fig:doublewell}(c)). As shown in \cref{fig:error}(left), Ehrenfest predictions deviate up to 450\% at the highest coupling, corresponding to a dramatic $7 \times$ rate enhancement (\cref{fig:ratesummary}(a)). In contrast, MASH with a classical cavity mode reduces the errors to 250\% (\cref{fig:error}(left))---a significant improvement, although still insufficient for accurate predictions (\cref{fig:ratesummary}(b)).

Introducing a quantum cavity mode further improves accuracy for both Ehrenfest and MASH, denoted as Ehrenfest+q and MASH+q, respectively. As shown in \cref{fig:ratesummary}(c,d), these methods produce comparable lineshapes, with errors below 50\% (\cref{fig:error}(left)). MASH+q, however, outperforms Ehrenfest+q in the predicted $k_\mathrm{res}/k_0$ and the absolute value of the rate on resonance (\cref{fig:error}(right)), where Ehrenfest+q predicts rates nearly twice as large. Notably, MASH with a classical cavity mode even outperforms Ehrenfest+q at weak coupling. A further key advantage of MASH lies in its consistency without the polaron transform. As shown in \cref{fig:nopolaron} of \cref{sec:moreresults}, MASH produces nearly identical results with or without the polaron transform, whereas Ehrenfest dynamics fails to predict resonant enhancement without it. This highlights the importance of the polaron transform in Ehrenfest dynamics and demonstrates MASH's reliability when using fewer Fock states. 

MASH+q achieves high overall accuracy, but significant errors—up to 50\%—persist at low coupling strength, as shown in \cref{fig:error}(a). 
These errors may stem from unphysical hopping between uncoupled states, a notable limitation known as size-inconsistency in the current version of multi-state MASH.\citep{runeson2023multi} 
This issue is illustrated in \cref{fig:mashqproblem}(a) and becomes apparent when simulating dynamics of the current model at zero coupling strength $(\eta_\mathrm{c}=0.0)$.
For instance, we see in \cref{fig:mashqproblem}(b) that MASH with quantum cavity mode allows transitions to excited states by increasing the photon number, even though the molecular system is uncoupled from the cavity mode. 
This behavior then affects the population dynamics, where MASH and MASH+q approach equilibrium at different rates when $\eta_\mathrm{c}=0.0$ (\cref{fig:mashqproblem}(c)). 
Notably, however, the two methods agree at short times, demonstrating how such dynamical errors are mitigated over the timescales in which rate calculations are typically performed. 
Nevertheless, to address this issue, we implement a threshold-based approach ($\epsilon$-MASH) as suggested in Ref.~\citenum{runeson2023multi}. Here, hops are rejected if the scalar nonadiabatic coupling $d_{an} = \sum_{j=1} p_j d^j_{an}$ falls below a threshold. By selecting $\epsilon = 7.5 \times 10^{-4}$, $\epsilon$-MASH ensures that the long-time decay of the quantum cavity mode aligns with the classical cavity mode in MASH at zero coupling, though minor discrepancies emerge at short times as a trade-off. 


\section{Conclusion}

In conclusion, we have presented a strategy based on the
mapping approach to surface hopping (MASH) and the use of
quantum cavity modes, to improve accuracy in rates calculated in models of vibrational polaritonic chemistry. 
The combination of MASH and a quantum cavity mode yields the most accurate rate predictions, agreeing more closely with the benchmark HEOM results. 
The use of the polaron transform further ensures computational efficiency and enable the calculation of consistent results when fewer Fock states are used. 
However, the observed discrepancies when uncoupled to the cavity highlight the need for further refinement of MASH to achieve consistent accuracy across both quantum and classical regimes. 

Future efforts will be directed to modifying the MASH framework for more robust simulations of polaritonic effects in complex molecular systems. While $\epsilon$-MASH addresses uncoupled hopping by tuning a threshold condition, this approach relies on trial-and-error to balance short- and long-time accuracy, making it an imperfect solution. An alternative multi-state MASH proposed in Ref.~\citenum{lawrence2024size} eliminates hopping between uncoupled states but performs poorly when many strongly coupled states are present, as seen in its reduced accuracy for the FMO model compared to the current multi-state MASH. Future work could explore applying this alternative multi-state MASH to the present problem to assess whether it improves accuracy, particularly at zero coupling.

We also plan to extend this work to the collective regime by explicitly simulating an ensemble of
dynamically evolving molecules. Our aim is to understand collective effects beyond the mean-field approximation used in previous works.\citep{fowler2022efficient,fowler2024mean,lindoy2024investigating}
The polaron transformation presents a particularly promising approach in this regime, as it enables rapid eigenspectrum convergence even as coupling strengths are enhanced by several orders of magnitude. 
This computational advantage should allow us to directly investigate how reaction rates at strong coupling scale with the number of molecules present in the system.

\section{Acknowledgements}
MRH acknowledges support by a grant from the Simons Foundation (Grant 839534, MET) and the NYU IT High Performance Computing resources, services, and staff expertise. AM acknowledges support from ACCESS (Advanced Cyberinfrastructure Coordination Ecosystem: Services \& Support) through allocation: PHY230021 which is supported by National Science Foundation grants \#2138259, \#2138286, \#2138307, \#2137603, and \#2138296. DRR acknowledges support from NSF CHE-2245592. MRH and AM acknowledge Pengfei Huo and Jeremy Richardson for useful discussions regarding estimating rates from direct population dynamics. MRH and AM also acknowledge Haimi Nguyen for pointing out initial inconsistencies in our code implementation of MASH.

\appendix

\section{Model Parameters}
\label{sec:modelparam}

\begin{table}
    \centering
    \begin{tabular}{cc}
    \hline \hline 
         Parameters& Value\\ \hline
         Mass $M$& 1.0 a.u.\\
         Barrier frequency $\omega_\mathrm{b}$& 1000 cm$^\mathrm{-1}$\\
         Energy barrier $E_\mathrm{b}$&2250 cm$^{-1}$ \\
         Solvent bath frequency $\gamma_\mathrm{s}$ & 200 cm$^{-1}$ \\
         Solvent reorganization energy $\lambda_\mathrm{s}$ & $\frac{1}{2} \Gamma_\mathrm{s} \gamma_\mathrm{s}$ \\
         Solvent friction $\Gamma_\mathrm{s}$ & $0.1 \omega_\mathrm{b}$ \\
         Cavity bath frequency $\gamma_\mathrm{c}$ & 1000 cm$^{-1}$ \\
         Cacity reorganization energy $\lambda_\mathrm{c}$ &  $\frac{\left(\omega_\mathrm{c}^2+\gamma_\mathrm{c}^2\right)\left(1-e^{-\beta \omega_\mathrm{c}} \right)}{2 \tau_\mathrm{c}\gamma_\mathrm{c}}$   \\ 
         Cavity lifetime & 2000 fs \\
         Temperature $T$ & 300 K
    \end{tabular}
    \caption{Fixed parameters for the model, adapted from previous works.\citep{lindoy2023quantum} Note that we sweep the cavity mode frequency $\omega_\mathrm{c}$ and coupling strength $\eta_\mathrm{c}$.}
    \label{tab:params}
\end{table}

Working in atomic units, we set $\hbar = 1.0$ and the molecule and cavity mode parameters are summarized in \cref{tab:params}. With these parameters, the transition frequency between the ground and excited state is $\omega_0 \approx 1190$ cm$^{-1}$. Since we are interested in the resonance effect, the range in which we sweep the cavity mode frequency is centered around $\omega_0$, i.e., $\omega_\mathrm{c} \in [690,1690]$ cm$^{-1}$. The coupling strengths are chosen to be $\eta_\mathrm{c} \in [6.25 \times 10^{-4}, 1.25 \times 10^{-3}, 2.5 \times 10^{-3}]$ a.u., which spanned the range chosen in previous work.\citep{hu2023resonance}

For all baths, the masses are set to unity, i.e., $m_i=1$ and $\tilde{m}_i=1$.
The coupling constants and frequencies of the solvent bath are specified by the spectral density 
\begin{equation}
J_\mathrm{S}(\omega) = \frac{\pi}{2} \sum_{i=1}^\infty \frac{g_i^2}{\omega_i} \delta(\omega-\omega_i) \,,
\end{equation}
which takes the Drude-Lorenz form,   
\begin{equation}
J_S(\omega) = \frac{2 \lambda_\mathrm{s} \gamma_\mathrm{s} \omega}{\omega^2+\gamma_\mathrm{s}^2}, 
\end{equation}
where $\lambda_\mathrm{s}$ and $\gamma_\mathrm{s}$ are the solvent reorganization and characteristic frequency, respectively. The Drude-Lorenz spectral density models a bath with exponential memory; dissipation emerges through a friction kernel $\Gamma_\mathrm{s}(t) = 2 \lambda_\mathrm{s} e^{-t \gamma_\mathrm{s}}$ that decays with a time constant $\tau_\mathrm{s} = \gamma_\mathrm{s}^{-1}$. Note that $\gamma_\mathrm{s}$ is chosen within the typical range of collective vibrational relaxation times at room temperatures. For the reorganization energy, we set its value based on the static friction defined as $\Gamma_\mathrm{s} := \int_0^\infty \Gamma_\mathrm{s}(t) = \frac{2 \lambda_\mathrm{s}}{\gamma_\mathrm{s}}$, and thus
\begin{equation}
\lambda_\mathrm{s} = \frac{1}{2} \gamma_\mathrm{s} \Gamma_\mathrm{s} \,.
\end{equation}
Following previous work,\citep{lindoy2023quantum} we set the friction constant to be $\Gamma_\mathrm{s} = 0.1 \omega_\mathrm{b}$.

The cavity loss bath follows from similar considerations. Defining it as
\begin{equation}
J_\mathrm{c}(\omega) = \frac{\pi}{2} \sum_{i=1}^\infty \frac{\tilde{g}_i^2}{\tilde{\omega}_i} \delta(\omega-\tilde{\omega}_i) \,,
\end{equation}
we again choose a Drude-Lorenz spectral density written as 
\begin{equation}
J_\mathrm{c}(\omega) = \frac{2 \lambda_\mathrm{c} \gamma_\mathrm{c} \omega}{\omega^2 +\gamma_\mathrm{c}^2} \,.
\end{equation}
The reorganization energy is determined from the cavity lifetime $\tau_\mathrm{c}$. In turn, the lifetime is obtained from applying Fermi's golden rule to compute the rate of single-photon absorption by the bath at thermal equilibrium.\citep{ying2023resonance} The result is
\begin{equation}
\tau_\mathrm{c}^{-1} = \frac{ J_S(\omega_\mathrm{c})}{\omega_\mathrm{c}(1-e^{-\beta \omega_\mathrm{c}})} \label{eq:fgrlifetime} \,,
\end{equation}
where $\beta = \frac{1}{k_\mathrm{B} T}$. From \cref{eq:fgrlifetime}, we obtain the reorganization energy as 
\begin{equation}
\lambda_\mathrm{s} = \frac{\left(\omega_\mathrm{c}^2+\gamma_\mathrm{c}^2\right)\left(1-e^{-\beta \omega_\mathrm{c}} \right)}{2 \tau_\mathrm{c}\gamma_\mathrm{c}} \,.
\end{equation}
The cavity characteristic frequency $\gamma_\mathrm{c}$ is typically chosen to ensure the bath operates in the Markovian regime, where the bath's friction acts instantaneously on the cavity mode without memory effects. 
To understand this Markovian behavior, we can treat the cavity mode and its associated bath as a single effective bath,\citep{ying2023resonance} which can be done by performing a normal mode transformation to write $\hat{H}_\mathrm{p}+\hat{H}_\mathrm{C} = \hat{H}_\mathrm{eff}$ where
\begin{equation}
\hat{H}_\mathrm{eff} = \sum_{i} \frac{\hat{p}_i^{\prime 2}}{2 }+ \frac{1}{2} \omega_i^2 \left(\hat{q}_i^\prime-\frac{g_i^\prime}{{\omega^\prime}_i^2} \hat{R} \right)^2 \,.
\end{equation}
The effective bath approach allows us to express the system as a molecule surrounded by a single composite harmonic bath. The coupling constants for this bath are specified by the effective spectral density
\begin{gather}
J_\mathrm{eff}(\omega) = \frac{2 \eta_\mathrm{c}^2 \omega_\mathrm{c}^3 J_\mathrm{c}(\omega)}{\left[\omega_\mathrm{c}^2 -\omega^2+\tilde{R}(\omega) \right]^2+\left[J_\mathrm{c}(\omega)\right]^2} \,,
\\
\tilde{R}(\omega) = \frac{\omega J_\mathrm{c}(\omega)}{\gamma_\mathrm{c}} \,. \label{eq:Rw}
\end{gather}
Note that \cref{eq:Rw} is unique to the Drude-Lorenz form of the cavity loss bath; see Ref.~\citenum{ying2023resonance} for a more general expression. 
In theory, we can tune to the Markovian limit by taking $\gamma_\mathrm{c} \to \infty$ such that $2 \lambda_\mathrm{c}/\gamma_\mathrm{c} = \Gamma_\mathrm{c}$ is finite. This implies $J_\mathrm{c}(\omega) \to \Gamma_\mathrm{c} \omega$, $\tilde{R}(\omega) \to 0$, and the spectral density becomes
\begin{equation}
J_\mathrm{eff}(\omega) \approx \frac{2 \eta_\mathrm{c}^2 \omega_\mathrm{c}^3  \Gamma_\mathrm{c} \omega }{\left(\omega_\mathrm{c}^2 -\omega^2 \right)^2+\Gamma_\mathrm{c}^2 \omega^2} \,.
\end{equation}
However, extremely high $\gamma_\mathrm{c}$ restricts the simulation timestep, making this limit infeasible for simulations.
Alternatively, we can take the limit where $\gamma_\mathrm{c}$ is sufficiently larger compared to the static friction constant of the cavity, i.e.,  $\Gamma_\mathrm{c}/\gamma_\mathrm{c} \to 0$. 
When restoring dimensions of the frequency with $\Gamma_\mathrm{c}$, we express $J_\mathrm{c}(\tilde{\omega}) = \tilde{\omega}/(1+\Gamma_\mathrm{c} \tilde{\omega}/\gamma_\mathrm{c}) \to \tilde{\omega} = \Gamma_\mathrm{c} \omega$ and $\tilde{R}(\tilde{\omega}) = \Gamma_\mathrm{c} \tilde{\omega} J(\tilde{\omega})/\gamma_\mathrm{c} \to 0$, namely the same result as the conventional Markovian limit. 
Given the current set of parameters, choosing $\gamma_\mathrm{c}= 1000$ cm$^{-1}$ is sufficient to obtain this limit with $\Gamma_\mathrm{c}/\gamma_\mathrm{c} \approx 6 \times 10^{-3}$. 


\label{sec:compdetails}



\section{Model Discretization} 
\label{sec:matelems}

To discretize the model, we begin with discretizing the reaction coordinate Hamiltonian $\hat{H}_\mathrm{R}$ by transforming the operator into a compact matrix defined in a truncated subspace of vibrational eigenstates of the molecule. 
This representation is obtained by starting from a position basis and adopting a discrete variable representation (DVR) where the 1D space over the reaction coordinate $R$ is discretized at a range $[-R_\mathrm{max},R_\mathrm{max}]$, where $R_\mathrm{max} = 100$ a.u., with the number of gridpoints $n_R = 2001$. The matrix elements of the kinetic energy operator become
\begin{gather}
(\mathbf{T}_\mathrm{R})_{ij} = \begin{cases} 
\frac{\pi^2}{6 M \Delta R^2} \left(1 + \frac{2}{N^2}\right)  & \text{if } i = j \\
\frac{\pi^2}{M N^2 \Delta R^2} \frac{(-1)^{j-i}}{\sin^2\left(\frac{\pi (j-i)}{N}\right)} & \text{if } i \neq j
\end{cases} \,.
\end{gather}
The matrix form of the coordinate operator $\hat{R}$ is diagonal and thus the matrix elements of the potential energy operator is also diagonal, 
\begin{gather}
(\mathbf{V}_\mathrm{R})_{ij} = 
\left[\frac{M^2 \omega_\mathrm{b}^4}{ 16 E_\mathrm{b}} \cdot R_i^4   -\frac{1}{2} M \omega_\mathrm{b}^2 \cdot R_i^2\right]  \delta_{ij} \,,
\end{gather}
where $\delta_{ij}$ is a Kronecker delta.  The discretized vibrational eigenstates, each of which is denoted by a vector $\mathbf{v}_\nu$, are obtained from diagonalizing the Hamiltonian matrix $\mathbf{H}_\mathrm{R} = \mathbf{T}_\mathrm{R} +\mathbf{V}_\mathrm{R}$, i.e., $\mathbf{H}_\mathrm{R} \mathbf{v}_\nu = E_\nu \mathbf{v}_\nu$. To work in a truncated subspace composed of the first four eigenstates, we construct an $n_R \times 4$ projection matrix $\mathbf{P} = \left[\mathbf{I}_{4 \times 4} \  \mathbf{0} \right]^T$ to obtain the transformation $\mathbf{P}^T\mathbf{H}_\mathrm{R} \mathbf{P} \to \mathbf{H}_\mathrm{R}$. 
From here, we can follow the diabatization outlined in \cref{sec:model} to obtain the final matrix
\begin{equation}
\mathbf{H}_\mathrm{R} = 
\begin{pmatrix}
\bar{E}_0 & \Delta_0 & 0 & 0 \\
\Delta_0 & \bar{E}_0 & 0 & 0 \\
0 & 0 & \bar{E}_1 & \Delta_1 \\
0 & 0 & \Delta_1 & \bar{E}_1
\end{pmatrix} \,. \label{eq:finalHr}
\end{equation}
Here, $\bar{E}_0 = \frac{1}{2}(E_0+E_1)$ and $\bar{E}_1 = \frac{1}{2}(E_2+E_1)$ represent the energies of the ground and excited states, respectively, and $\Delta_0 = \frac{1}{2}(E_1-E_0)$ and $\Delta_1 = \frac{1}{2}(E_3-E_2)$ are the diabatic couplings. With current parameters, $\bar{E}_0 = 1190$ cm$^{-1}$ and we expect resonance enhancement to occur at this value. 

When treating the cavity mode classically, \cref{eq:finalHr} completely defines the quantum system of the MQC simulation. If the cavity mode is quantized, we need to adopt further discretization. As explained in \cref{sec:qph}, this can be achieved with a Fock basis, truncated to $N_\mathrm{p}$-many states. The result of adopting a Fock basis is the following matrix elements for the free-photon Hamiltonian 
\begin{equation}
(\mathbf{H}_\mathrm{p})_{NM} =  \omega_\mathrm{c}\left(N+\frac{1}{2}\right)  \delta_{NM} \,,
\end{equation}
where $\delta_{NM}$ is a Kronecker delta. 
The creation and annihilation operators in matrix form are 
\begin{equation}
\left(\mathbf{a}^\dagger \right)_{NM} = \begin{cases}
\sqrt{N} & \text{if} \ N=M-1
\\
0 & \text{otherwise}
\end{cases}; \quad \mathbf{a} = \left(\mathbf{a}^\dagger\right)^T \,,
\end{equation}
and thus the matrix form of the PF Hamiltonian is 
\begin{equation}
\mathbf{H}_\mathrm{PF} = \mathbf{I}_\mathrm{p} \otimes \mathbf{H}_\mathrm{R}+\mathbf{H}_\mathrm{p} \otimes \mathbf{I}_\mathrm{R}+\mathbf{H}_\mathrm{int} \,, \label{eq:pfhamilmatrix}
\end{equation}
where $\otimes$ is the Kronecker product, $\mathbf{I}_\mathrm{p}$ and $\mathbf{I}_\mathrm{R}$  are identity matrices of size $N_\mathrm{p} \times N_\mathrm{p}$ and $ 4 \times 4 $, respectively. The interaction term can be written as 
\begin{equation}
\mathbf{H}_\mathrm{int} = \omega_\mathrm{c} \eta_\mathrm{c} \mathbf{R} \otimes \left(\mathbf{a} +\mathbf{a}^\dagger\right)+ \omega_\mathrm{c} \eta_\mathrm{c}^2 \mathbf{R}^2 \otimes \mathbf{I}_\mathrm{p} \,,
\end{equation}
where $\mathbf{R}$ is the matrix form of the reaction coordinate operator in the truncated vibrational eigenbasis. \Cref{eq:pfhamilmatrix} sets the quantum system when employing the quantum cavity mode. Note that for all calculations we set $N_\mathrm{p}=2$, i.e., the single-photon exictation subspace. 

For the polaron-transform version of \cref{eq:pfhamilmatrix}, we first transform the reaction coordinate Hamiltonian to its MH basis, i.e., $ \mathbf{U}^T_\mathrm{MH} \mathbf{H}_\mathrm{R} \mathbf{U}_\mathrm{MH} \to \mathbf{H}_\mathrm{R}$ where the transformation $\mathbf{U}_\mathrm{MH} = [\mathbf{u}_1, \mathbf{u}_2, \mathbf{u}_3, \mathbf{u}_4]$ is composed of eigenvectors of the reaction coordinate operator, i.e., $ \mathbf{R} \mathbf{u}_\nu = R_\nu \mathbf{u}_\nu$. After applying the polaron transform, we then obtain the following expression
\begin{equation}
\mathbf{H}_\mathrm{PF}^\prime = \mathbf{H}_\mathrm{R}^\prime +\mathbf{H}_\mathrm{P} \otimes \mathbf{I}_\mathrm{M} \,, \label{eq:pfhamiltonianpolaron}
\end{equation}
where the matrix elements of the reaction coordinate Hamiltonian after the polaron transformation is 
\begin{equation}
\left(\mathbf{H}_\mathrm{R}^{\prime}\right)_{\nu \mu; NM} = \left( \mathbf{H}_\mathrm{R} \right)_{\nu \mu} S_{\nu \mu; NM} \,,
\end{equation}
\begin{equation}
S_{\nu\mu; NM} = \int \mathrm{d} q \ \psi_N\left(q-q_\nu\right) \psi_M\left(q-q_\mu\right) \,,
\end{equation}
where  $q_\nu = -\eta_\mathrm{c} R_\nu \sqrt{\tfrac{2}{\omega_\mathrm{c}}}$ and $S_{NM; \nu \mu}$ is an overlap integral of the photon wavefunctions,
\begin{equation}
\psi_N(r) = \frac{1}{\sqrt{2^N N!}} \left(\frac{\omega_\mathrm{c}}{\pi}\right)^{1/4} e^{-\frac{\omega_\mathrm{c} r^2}{2}} h_N(\sqrt{\omega_\mathrm{c}} r) \,.
\end{equation}
Here, $h_N(r)$ is the $N$-th order Hermite polynomial. The overlap integral is computed using the trapezoidal rule with the same DVR gridpoints. 
Note that the polaron transform also affects the matrix form of the photon displacement operator $\mathbf{Q}_\mathrm{c}$ by translating it as $\mathbf{Q}_\mathrm{c}^\prime = \mathbf{I}_\mathrm{M} \otimes \mathbf{Q}_\mathrm{c}+ \eta_\mathrm{c} \mathbf{R}^\prime \otimes \mathbf{I}_\mathrm{p} \sqrt{\frac{2}{\omega_\mathrm{c}}}$. All the results from MQC simulations with quantum cavity mode in \cref{sec:results} are obtained using the polaron transform. 

We discretize the solvent and cavity loss baths such that every mode carries an equal amount of the total reorganization energy. 
Following Ref.~\citenum{walters2017direct}, we can relate an arbitrary spectral density $J(\omega)$ with the reorganization energy using the following integral
\begin{gather}
\lambda = \int_0^\infty \dif \omega \ \frac{J (\omega)}{\pi \omega} =  \sum_{i=1}^{\infty} \frac{g_i^2}{2 \omega_i^2} \,.
\end{gather}
Next, we choose a maximum frequency $\omega_\mathrm{max}$ so that we approximate the reorganization energy as
\begin{equation}
\tilde{\lambda} =  \int_0^{\omega_\mathrm{max}} \dif \omega \ \frac{J(\omega)}{\pi \omega}  = F(\omega_\mathrm{max}) \,.
\end{equation}
To assign an equal fraction of the reorganization energy for every $j$-th mode, we partition $\tilde{\lambda}$ into $N_\mathrm{B}$-many pieces and set $\frac{g_i^2}{2 \omega_i^2} = \frac{\tilde{\lambda}}{N_\mathrm{B}}$. 
The coupling constant can then be written as 
\begin{equation}
g_i =  \omega_i \sqrt{\frac{2 \tilde{\lambda}}{N_\mathrm{B}}} \,,
\end{equation}
and the frequency is chosen so that 
\begin{gather}
\omega_i = \argmin_{\omega \in [0,\omega_\mathrm{max}]} \left[F(\omega) - \left(\tfrac{i-1/2}{N_\mathrm{B}}\right)\tilde{\lambda}\right] \,.
\end{gather}
For the solvent and cavity bath,s we can write the cumulative function as $F_\mathrm{S}(\omega)=\frac{2 \lambda_\mathrm{s}}{\pi} \tan^{-1}\left( \frac{\omega_\mathrm{max}^\mathrm{S}}{\gamma_\mathrm{s}}\right)$ and $F_\mathrm{c}(\omega)=\frac{2 \lambda_\mathrm{c}}{\pi} \tan^{-1}\left( \frac{\omega_\mathrm{max}^\mathrm{c}}{\gamma_\mathrm{c}}\right)$, respectively. 
We choose  $\omega_\mathrm{max}^\mathrm{s} = 10 \gamma_\mathrm{s}$ and $\omega_\mathrm{max}^\mathrm{c} = 3 \gamma_\mathrm{c}$, respectively, and $N_\mathrm{B}=200$ for all baths.
When treating the cavity mode classically, we opt to use the effective spectral density with the cumulative function $F_\mathrm{eff}(\omega)$ computed numerically using $1000N_\mathrm{B}$ gridpoints to approximate the frequency integral. 

\section{Simulation Details}

The setup of a MQC simulation begins with setting up its initial conditions, propagating the dynamics with discretized equations of motion, and computing observables based on the rules of the MQC method. Here, we outline how that is done in Ehrenfest and MASH, respectively. In all subsequent discussions, the matrix form of the energy operator $\hat{E}(\mathbf{q})$ when using a classical cavity mode is
\begin{gather}
E_{\nu \mu}(\mathbf{q}) = (\mathbf{H}_\mathrm{R})_{\nu \mu}+(\mathbf{H}_\mathrm{S}(\mathbf{q}))_{\nu \mu} +(\mathbf{H}_\mathrm{eff}(\mathbf{q}))_{\nu \mu} \,,
\\
\mathbf{H}_\mathrm{S}(\mathbf{q}) = \sum_i \left[\frac{1}{2} \tilde{m} \omega_i^2 q_i^2 \mathbf{I} - g_i q_i \mathbf{R}+\frac{1}{2} \frac{g_i^2}{\tilde{m} \omega_i^2} \mathbf{R}^2 \right] \,,
\\
\mathbf{H}_\mathrm{eff}(\mathbf{q}) = \sum_i \left[\frac{1}{2} \tilde{m} \omega_i^{\prime 2} q_i^{\prime 2} \mathbf{I} - g_i q_i^\prime \mathbf{R}+\frac{1}{2} \frac{g_i^{\prime 2}}{\tilde{m} \omega_i^{\prime 2}} \mathbf{R}^2 \right]  \,,
\end{gather}
where $\mathbf{q} = [ q_1, \ldots, q_{N_\mathrm{B}}, q_1^\prime, \ldots, q_{N_\mathrm{B}}^\prime ]^T$ is the solvent and effective bath coordinates as a single vector and we set the masses for all classical DOFs to $\tilde{m}=1$.  In a quantum cavity mode, we have 
\begin{gather}
E_{\nu \mu}(\mathbf{q}) = (\mathbf{H}_\mathrm{PF})_{\nu \mu}+(\mathbf{H}_\mathrm{S}(\mathbf{q}))_{\nu \mu} +(\mathbf{H}_\mathrm{C}(\mathbf{q}))_{\nu \mu}  \,,
\\
\mathbf{H}_\mathrm{S}(\mathbf{q}) = \sum_i \left[\frac{1}{2} \tilde{m} \omega_i^2 q_i^2 \mathbf{I} - g_i q_i \mathbf{R}+\frac{1}{2} \frac{g_i^2}{\tilde{m} \omega_i^2} \mathbf{R}^2 \right]  \,,
\\
\mathbf{H}_\mathrm{C}(\mathbf{q}) = \sum_i \left[\frac{1}{2} \tilde{m} \tilde{\omega}_i^{2} r_i^{2} \mathbf{I} - \tilde{g}_i r_i \mathbf{Q}_\mathrm{c}+\frac{1}{2} \frac{\tilde{g}_i^{2}}{\tilde{m} \tilde{\omega}_i^{\prime 2}} \mathbf{Q}_\mathrm{c}^2 \right]  \,,
\end{gather}
where $\mathbf{q} = [ q_1, \ldots, q_{N_\mathrm{B}}, r_1, \ldots, r_{N_\mathrm{B}}]^T$ is the solvent and cavity bath coordinates as a single vector. When using the polaron transform version, we replace $\mathbf{H}_\mathrm{PF}$ and $\mathbf{Q}_\mathrm{c}$ with their polaron transform versions, $\mathbf{H}^\prime_\mathrm{PF}$ and $\mathbf{Q}_\mathrm{c}^\prime$, respectively. 


\subsection{Ehrenfest}

For Ehrenfest simulations, the molecule is initialized in the left ground state $| \nu_\mathrm{L} \rangle$. When the cavity mode is classical, this implies that the initial coefficient is $c_\nu^0=1$ if $\nu = 0$ and zero otherwise. When the cavity mode is quantized, we initialize the photonic state thermally by selecting an $N$-photon state $| N \rangle$ with probability $p_N = e^{- \beta E_N}/Z$ where $Z  =\sum_{N=1}^{N_\mathrm{p}} e^{-\beta E_N}$ and  $E_N = \omega_\mathrm{c} (N+1/2)$. Thus, the initial vector $c_{\nu;N}^0 = 1$ if $\nu = 1$ and $N \sim p_N$. Note that the ground state $p_0$ is often chosen with nearly $99\%$ probability at room temperature. 

The classical degrees of freedom, $(\mathbf{q},\mathbf{p})$ in vector notation, are initialized through their thermal Wigner distributions, each of which is a normal distribution for harmonic DOFs. Setting the mass $m=1$, the initial classical momenta and positions are 
\begin{equation}
\mathbf{p} \sim \mathcal{N}(0,\bm{\Sigma}_p), \quad \mathbf{q} \sim \mathcal{N}(\mathbf{q}_0,\bm{\Sigma}_q) \,,
\end{equation}
where $(\bm{\Sigma}_p)_{ij} = \sqrt{\frac{\omega_i}{2 \tanh( \beta \hbar \omega_i/2)}} \delta_{ij}$ and $(\bm{\Sigma}_q)_{ij} = \sqrt{\frac{1}{2 \omega_i \tanh( \beta \hbar \omega_i/2)}} \delta_{ij}$ and $(\mathbf{q}_0)_i = g_i \langle \hat{R} \rangle_0/\omega_i^2$ where $\langle \hat{R} \rangle_0 = \sum_{\nu \mu} c_\nu^{0*} R_{\nu \mu} c_\mu^0$ is the initial average value of the coordinate operator, which should be concentrated in the left well.

Dynamics is propagated in the diabatic basis by discretizing the TDSE with a 4th-order Runge-Kutta (RK4) scheme. Given a timestep $\delta t$, the RK4 scheme is
\begin{gather}
c_\nu \leftarrow c_\nu+\frac{\delta t }{6}\left(k^{(1)}_\nu+k^{(2)}_\nu+k^{(3)}_\nu+k^{(4)}_\nu\right) \,,
\\
k_\nu^{(1)}  = -\mathrm{i} E_{\nu \mu}(\mathbf{q}) c_\mu \,,
\\
k_\nu^{(2)} = -\mathrm{i} E_{\nu \mu}(\mathbf{q}) \left(c_\mu+\tfrac{\delta t}{2}k_\mu^{(1)}\right) \,,
\\
k_\nu^{(3)} = -\mathrm{i} E_{\nu \mu}(\mathbf{q}) \left(c_\mu+\tfrac{\delta t}{2}k_\mu^{(2)}\right) \,,
\\
k_\nu^{(4)} =  -\mathrm{i} E_{\nu \mu}(\mathbf{q}) \left(c_\mu+\delta t k_\mu^{(3)}\right) \,,
\end{gather}
We combine RK4 scheme for the TDSE with the velocity Verlet scheme for the classical system to have the following update rule at given timestep $\Delta t$ 
\begin{gather}
c_\nu \leftarrow \mathrm{RK4}(\{c_\nu\},\mathbf{q}; \Delta t/2) \,,
\\
\mathbf{p} \leftarrow \mathbf{p} -\sum_{\nu,\mu} c_\nu^* c_\mu \nabla_{\mathbf{q}} E_{\nu \mu}(\mathbf{q})\Delta t / 2  \,,
\\
\mathbf{q} \leftarrow \mathbf{q}+(\mathbf{p} / \tilde{m}) \Delta t  \,,
\\
\mathbf{p} \leftarrow \mathbf{p} -\sum_{\nu,\mu} c_\nu^* c_\mu \nabla_{\mathbf{q}} E_{\nu \mu}(\mathbf{q})  \Delta t / 2 \,,
\\
c_\nu \leftarrow \mathrm{RK4}(\{c_\nu\}, \mathbf{q}; \Delta t/2) \,.
\end{gather}

As dynamics propagate over time, we calculate the reactant population by writing down a Heaviside operator where $h(R) = 1$ if $R < 0$ and zero otherwise. In the DVR representation it becomes $(\mathbf{h}_R)_{ij} = h(R_i) \delta_{ij}$. Afterwards, we transform it to the truncated subspace of vibrational eigenstates yielding  $\mathbf{P}^T(\mathbf{U}^T \mathbf{h}_R \mathbf{U})\mathbf{P} \to \mathbf{h}_R$. Constructing the density matrix as $\bm{\rho} = \mathbf{c} \otimes \mathbf{c}$, we then compute the reactant population as 
\begin{equation}
P_R = \mathrm{Tr} \left[ \mathbf{h}_R \bm{\rho} \right]  \,. \label{eq:prrho}
\end{equation}
To obtain statistically convergent result, we initiate $10^6$ trajectories for every set of model parameters.

\subsection{MASH}

In MASH, the initial state is chosen the same way as Ehrenfest. However, the values we subtitute into the initial vector are chosen stochastically. When we work with a classical cavity mode, we use focused sampling: 
\begin{equation}
c_\nu = \begin{cases}
\sqrt{\frac{1-\beta_N}{\alpha_N}} e^{2 \pi \mathrm{i} \phi_\nu } & \text{if} \ \nu = 0
\\
-\sqrt{\frac{\beta_N}{\alpha_N}} e^{2 \pi \mathrm{i} \phi_\nu} & \text{otherwise}
\end{cases} \,, \label{eq:cnu}
\end{equation}
where $\phi_\nu \in \mathrm{Unif}[0,1]$ and
\begin{equation}
\alpha_{N_\mathrm{s}} = \frac{N_\mathrm{s}-1}{-1+\sum_{n=1}^{N_\mathrm{s}} 1/n} \,, \quad \beta_{N_\mathrm{s}} = \frac{1-\alpha_{N_\mathrm{s}}}{N_\mathrm{s}} \,, \label{eq:mappingcoeff}
\end{equation}
where $N_\mathrm{s} = 4 N_\mathrm{p}$ is the total number of states and is the product of 4 vibrational and $N_\mathrm{p}$ states; note that other forms of sampling are also possible.\citep{runeson2024exciton}  
The coefficients in \cref{eq:mappingcoeff} are part of the multi-state mapping adopted in MASH, and are derived so that populations and coherences can be computed from any basis.\citep{runeson2023multi}
When using a quantum cavity mode, we replace $\nu \to (\nu,N)$ with $\nu=0$ and $N \sim p_N=e^{-\beta \omega_\mathrm{c}(N+1/2)}/Z$.  Since dynamics is propagated in an adiabatic basis, we transform the initial vector via a unitary matrix $\mathbf{W}$. For all simulations without using the polaron transform, $\mathbf{W}$ is constructed from diagonalizing the energy matrix $ \mathbf{E}(\mathbf{q})$, which yields $\mathbf{\tilde{U}}$, and thus $\mathbf{W} = \mathbf{\tilde{U}}$. With the polaron transform, we need to transform from the MH basis to the diabatic basis before applying $\mathbf{\tilde{U}}$, and thus $\mathbf{W} = \mathbf{\tilde{U}} \mathbf{U}_\mathrm{MH}$. 
Note the classical degrees of freedom are also initialized as in Ehrenfest but the initial vector is specified in \cref{eq:cnu}.

The dynamics provided by MASH must account for the continuous dynamics on the potential energy surface (PES) of an active $a$ and the impulse force responsible for switching between active states. To this end, we follow Ref.~\citenum{runeson2023multi} to implement the local diabatization method\citep{granucci2001direct} to propagate the quantum system, and the velocity Verlet scheme for the classical system as the dynamics proceed on active PES:
\begin{gather}
c_n \leftarrow \mathrm{e}^{-\mathrm{i} E_n(\mathbf{q}) \Delta t / 2} c_n\,, \label{eq:adprop}
\\
c_\nu = U_{\nu n}^T c_n, \quad U_{n \nu } E_{\nu \mu }(\mathbf{q}) = E_n U_{n \mu} \,, \label{eq:adprop1}
\\ 
\mathbf{p} \leftarrow \mathbf{p}-\nabla_\mathbf{q} E_a(\mathbf{q}) \Delta t / 2 \,, \\
\mathbf{q} \leftarrow \mathbf{q}+(\mathbf{p} / m) \Delta t \,, \\
\mathbf{p} \leftarrow \mathbf{p}-\nabla_\mathbf{q} E_a(\mathbf{q}) \Delta t / 2\,, \\
c_n = U_{n \nu} c_\nu, \quad U_{n \nu } E_{\nu \mu }(\mathbf{q}) = E_n U_{n \mu}\,,  \label{eq:adprop2} \\ 
c_n \leftarrow \mathrm{e}^{-\mathrm{i} E_n(\mathbf{q}) \Delta t / 2} c_n\,, \label{eq:adprop3}
\end{gather}
Note that \cref{eq:adprop,eq:adprop1,eq:adprop2,eq:adprop3} is a scheme to propagate the wavefunction coefficients in the adiabatic basis without explicitly computing nonadiabatic coupling vectors. 

\begin{figure*}
\includegraphics[width=\linewidth]{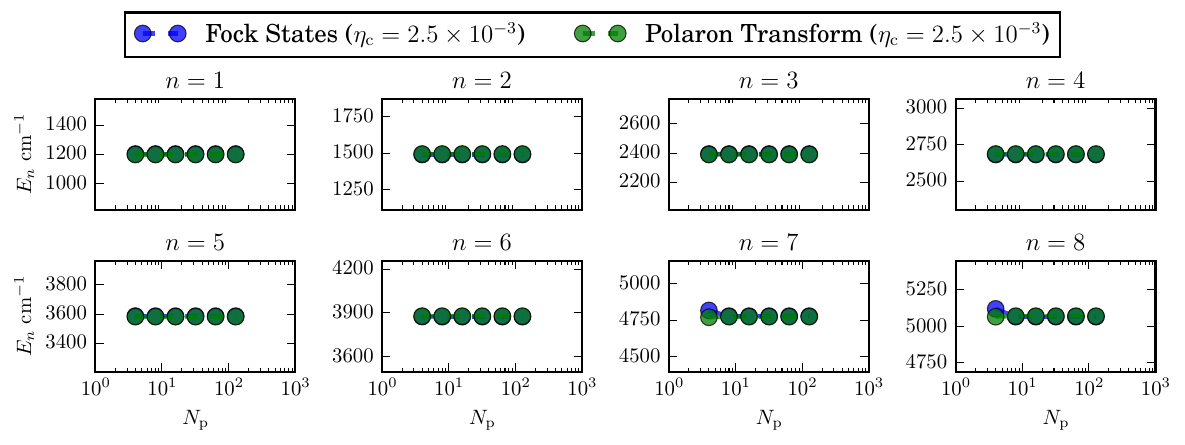}
\caption{Convergence of the energy in the first eight adiabatic states at $\eta_\mathrm{c} = 2.5 \times 10^{-3}$ a.u.}
\label{fig:loweta_eig}
\end{figure*}

\begin{figure*}
\includegraphics[width=\linewidth]{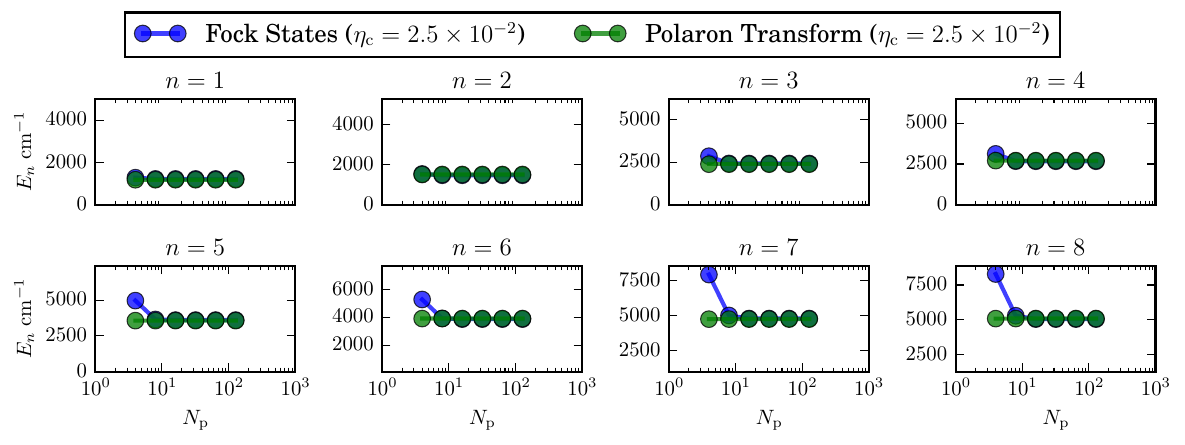}
\caption{Convergence of the energy in the first eight adiabatic states at $\eta_\mathrm{c} = 2.5 \times 10^{-2}$ a.u.}
\label{fig:mideta_eig}
\end{figure*}

\begin{figure*}
\includegraphics[width=\linewidth]{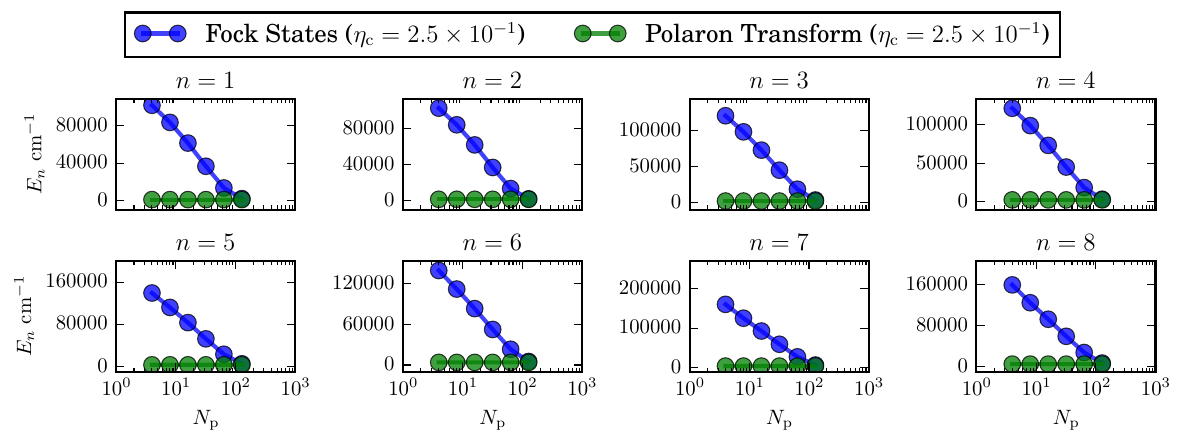}
\caption{Convergence of the energy in the first eight adiabatic states at $\eta_\mathrm{c} = 2.5 \times 10^{-1}$ a.u.}
\label{fig:higheta_eig}
\end{figure*}

\begin{figure*}
    \centering
    \includegraphics[width=0.685\linewidth]{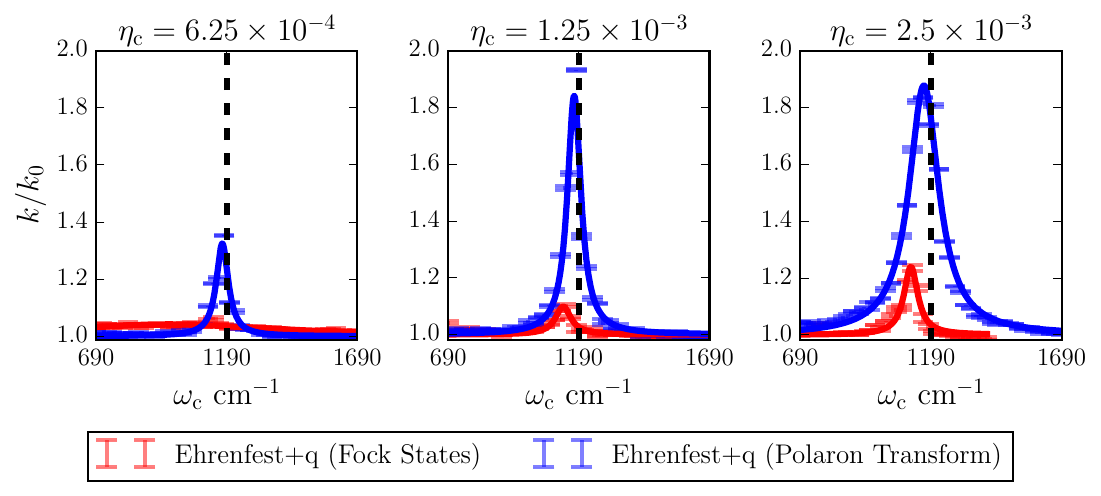}
    \includegraphics[width=0.685\linewidth]{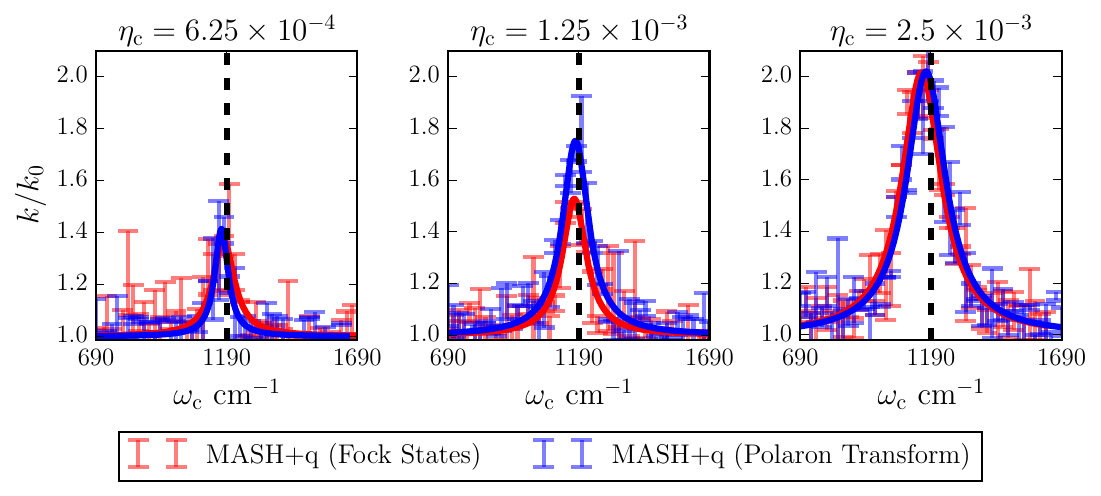}
    \caption{The rate profile over cavity-mode frequency $\omega_\mathrm{c}$ in (top) Ehrenfest+q and (bottom) MASH+q, with and without the polaron transform. The dashed vertical line corresponds to the resonant frequency at $\omega_0 = 1190 \ \mathrm{cm}^{-1}$. Error bars represent 99\% confidence intervals; see \cref{sec:compdetails} for details.}
    \label{fig:nopolaron}
\end{figure*}

When the instantaneous population of the active state satisfies $\rho_a = \rho_b$, where $b$ is another adiabatic state, the system must determine if switching between the states is physically feasible---specifically, whether there is enough kinetic energy to overcome the energy barrier $E_a - E_b$. To evaluate this, the simulation rolls back the new timestep and re-propagates the dynamics for half a timestep, $\Delta t/2$. At this intermediate point, the feasibility of hopping is tested by projecting the mass-weighted momenta, $\tilde{\mathbf{p}} = \mathbf{p}/\sqrt{\tilde{m}}$, onto the coupling vector $\bm{\delta}_{ab} = \bm{\nabla} (\rho_a - \rho_b)$. The explicit form of $\bm{\delta}_{ab}$ in terms of wavefunction coefficients and nonadiabatic coupling vectors is
\begin{equation}
\bm{\delta}_{ab} = \frac{1}{\sqrt{\tilde{m}}} \sum_{a'} \operatorname{Re}\left[c_{a'}^* \mathbf{d}_{a'a} c_a - c_{a'}^* \mathbf{d}_{a'b} c_b\right] \,.
\end{equation} 
Since the projection of the momenta onto the coupling vector is $\tilde{\mathbf{p}}^{||} = (\tilde{\mathbf{p}} \cdot \bm{\delta}_{ab}) / |\bm{\delta}_{ab}|$, the projected kinetic energy is then $|\tilde{\mathbf{p}}^{||}|^2 / 2$. If this energy is less than the barrier height, $E_a - E_b$, the system lacks sufficient kinetic energy to switch states, and the active state remains in $a$. The dynamics are then propagated for another $\Delta t/2$ on the current surface. If the kinetic energy is above the threshold, the magnitude of the projected momentum is rescaled as $|\tilde{\mathbf{p}}^{||}| \leftarrow \sqrt{|\tilde{\mathbf{p}}^{||}|^2 + 2(E_a - E_b)}$, while the orthogonal components remain unchanged. The dynamics are then further propagated for an additional $\Delta t/2$ on the new active surface as the state switches from $a$ to $b$. 
Note that this scheme approximates the bisection method proposed in the original paper\citep{runeson2023multi}, which we adopt to save computational time in the simulation.

To compute the reactant population we first transform the vector of coefficients, written now as $\mathbf{c}$, back to the diabatic basis $ \mathbf{W} \mathbf{c} \to \mathbf{c}$ with the density matrix computed as 
\begin{equation}
\bm{\rho} = \alpha_{N_\mathrm{s}} \left(\mathbf{c} \otimes \mathbf{c} \right)+\beta_{N_\mathrm{s}} \mathbf{I}_{N_\mathrm{s} \times N_\mathrm{s}} \,, \label{eq:rhomash}
\end{equation}
where $\mathbf{I}_{N_\mathrm{s} \times N_\mathrm{s}}$ is $N_\mathrm{s} \times N_\mathrm{s}$ identity matrix and we compute reactant population using \cref{eq:prrho}. 
See Appendices~B and C of Ref.~\citenum{runeson2023multi} for the derivation of \cref{eq:rhomash}. To obtain statistically convergent result, we initiate $10^6$ trajectories for every set of model parameters.

\section{Further Calculations Utilizing Polaron Transform}
\label{sec:moreresults}

We compare calculations using quantum cavity modes, with and without the polaron transform. First, we begin with \cref{fig:loweta_eig,fig:mideta_eig,fig:higheta_eig}, which highlight the convergence of adiabatic energies for the full Hamiltonian, including system-bath coupling. 
The bath degrees of freedom are initialized assuming the quantum subsystem sits at the $ |\nu_L, 0\rangle $ state.
At the lowest coupling strength (\cref{fig:loweta_eig}), convergence is faster with the polaron transform, which only requires $N_\mathrm{p}=2$ to reach convergence in comparison to $ N_\mathrm{p} > 10 $ without it. 
The advantages of the polaron transform become pronounced at higher coupling strengths, where errors exceed $O(10^3) \ \mathrm{cm}^{-1} $ for $ N_\mathrm{p} = 2 $ without the transform. 
At the highest coupling strength (\cref{fig:higheta_eig}), achieving convergence without the polaron transform requires $N_\mathrm{p} \gg 10^2$, while the polaron transform achieves accuracy with only $N_\mathrm{p} = 2$. This highlights the polaron transform's effectiveness in describing the quantum subsystem at strong coupling. 

In \cref{fig:nopolaron}, we see the advantages of polaron transform from the rate calculations at different coupling strengths. A notable discrepancy arises in the Ehrenfest method, where significant deviations occur, unlike MASH, which remains consistent across all coupling strengths. 

\nocite{*}
{\small \bibliography{achemso-demo}}

\providecommand{\latin}[1]{#1}
\makeatletter
\providecommand{\doi}
  {\begingroup\let\do\@makeother\dospecials
  \catcode`\{=1 \catcode`\}=2 \doi@aux}
\providecommand{\doi@aux}[1]{\endgroup\texttt{#1}}
\makeatother
\providecommand*\mcitethebibliography{\thebibliography}
\csname @ifundefined\endcsname{endmcitethebibliography}  {\let\endmcitethebibliography\endthebibliography}{}
\begin{mcitethebibliography}{43}
\providecommand*\natexlab[1]{#1}
\providecommand*\mciteSetBstSublistMode[1]{}
\providecommand*\mciteSetBstMaxWidthForm[2]{}
\providecommand*\mciteBstWouldAddEndPuncttrue
  {\def\EndOfBibitem{\unskip.}}
\providecommand*\mciteBstWouldAddEndPunctfalse
  {\let\EndOfBibitem\relax}
\providecommand*\mciteSetBstMidEndSepPunct[3]{}
\providecommand*\mciteSetBstSublistLabelBeginEnd[3]{}
\providecommand*\EndOfBibitem{}
\mciteSetBstSublistMode{f}
\mciteSetBstMaxWidthForm{subitem}{(\alph{mcitesubitemcount})}
\mciteSetBstSublistLabelBeginEnd
  {\mcitemaxwidthsubitemform\space}
  {\relax}
  {\relax}

\bibitem[Dunkelberger \latin{et~al.}(2022)Dunkelberger, Simpkins, Vurgaftman, and Owrutsky]{dunkelberger2022vibration}
Dunkelberger,~A.~D.; Simpkins,~B.~S.; Vurgaftman,~I.; Owrutsky,~J.~C. Vibration-cavity polariton chemistry and dynamics. \emph{Ann. Rev. Phys. Chem.} \textbf{2022}, \emph{73}, 429--451\relax
\mciteBstWouldAddEndPuncttrue
\mciteSetBstMidEndSepPunct{\mcitedefaultmidpunct}
{\mcitedefaultendpunct}{\mcitedefaultseppunct}\relax
\EndOfBibitem
\bibitem[Thomas \latin{et~al.}(2016)Thomas, George, Shalabney, Dryzhakov, Varma, Moran, Chervy, Zhong, Devaux, Genet, \latin{et~al.} others]{thomas2016ground}
Thomas,~A.; George,~J.; Shalabney,~A.; Dryzhakov,~M.; Varma,~S.~J.; Moran,~J.; Chervy,~T.; Zhong,~X.; Devaux,~E.; Genet,~C.; others Ground-state chemical reactivity under vibrational coupling to the vacuum electromagnetic field. \emph{Angew. Chem.} \textbf{2016}, \emph{128}, 11634--11638\relax
\mciteBstWouldAddEndPuncttrue
\mciteSetBstMidEndSepPunct{\mcitedefaultmidpunct}
{\mcitedefaultendpunct}{\mcitedefaultseppunct}\relax
\EndOfBibitem
\bibitem[Thomas \latin{et~al.}(2019)Thomas, Lethuillier-Karl, Nagarajan, Vergauwe, George, Chervy, Shalabney, Devaux, Genet, Moran, \latin{et~al.} others]{thomas2019tilting}
Thomas,~A.; Lethuillier-Karl,~L.; Nagarajan,~K.; Vergauwe,~R.~M.; George,~J.; Chervy,~T.; Shalabney,~A.; Devaux,~E.; Genet,~C.; Moran,~J.; others Tilting a ground-state reactivity landscape by vibrational strong coupling. \emph{Science} \textbf{2019}, \emph{363}, 615--619\relax
\mciteBstWouldAddEndPuncttrue
\mciteSetBstMidEndSepPunct{\mcitedefaultmidpunct}
{\mcitedefaultendpunct}{\mcitedefaultseppunct}\relax
\EndOfBibitem
\bibitem[Lather \latin{et~al.}(2019)Lather, Bhatt, Thomas, Ebbesen, and George]{lather2019cavity}
Lather,~J.; Bhatt,~P.; Thomas,~A.; Ebbesen,~T.~W.; George,~J. Cavity catalysis by cooperative vibrational strong coupling of reactant and solvent molecules. \emph{Angew. Chem. Int. Ed.} \textbf{2019}, \emph{58}, 10635--10638\relax
\mciteBstWouldAddEndPuncttrue
\mciteSetBstMidEndSepPunct{\mcitedefaultmidpunct}
{\mcitedefaultendpunct}{\mcitedefaultseppunct}\relax
\EndOfBibitem
\bibitem[Vergauwe \latin{et~al.}(2019)Vergauwe, Thomas, Nagarajan, Shalabney, George, Chervy, Seidel, Devaux, Torbeev, and Ebbesen]{vergauwe2019modification}
Vergauwe,~R.~M.; Thomas,~A.; Nagarajan,~K.; Shalabney,~A.; George,~J.; Chervy,~T.; Seidel,~M.; Devaux,~E.; Torbeev,~V.; Ebbesen,~T.~W. Modification of enzyme activity by vibrational strong coupling of water. \emph{Angew. Chem. Int. Ed.} \textbf{2019}, \emph{58}, 15324--15328\relax
\mciteBstWouldAddEndPuncttrue
\mciteSetBstMidEndSepPunct{\mcitedefaultmidpunct}
{\mcitedefaultendpunct}{\mcitedefaultseppunct}\relax
\EndOfBibitem
\bibitem[Lather and George(2020)Lather, and George]{lather2020improving}
Lather,~J.; George,~J. Improving enzyme catalytic efficiency by co-operative vibrational strong coupling of water. \emph{J. Phys. Chem. Lett.} \textbf{2020}, \emph{12}, 379--384\relax
\mciteBstWouldAddEndPuncttrue
\mciteSetBstMidEndSepPunct{\mcitedefaultmidpunct}
{\mcitedefaultendpunct}{\mcitedefaultseppunct}\relax
\EndOfBibitem
\bibitem[Hirai \latin{et~al.}(2020)Hirai, Takeda, Hutchison, and Uji-i]{hirai2020modulation}
Hirai,~K.; Takeda,~R.; Hutchison,~J.~A.; Uji-i,~H. Modulation of prins cyclization by vibrational strong coupling. \emph{Angew. Chem.} \textbf{2020}, \emph{132}, 5370--5373\relax
\mciteBstWouldAddEndPuncttrue
\mciteSetBstMidEndSepPunct{\mcitedefaultmidpunct}
{\mcitedefaultendpunct}{\mcitedefaultseppunct}\relax
\EndOfBibitem
\bibitem[Ahn \latin{et~al.}(2023)Ahn, Triana, Recabal, Herrera, and Simpkins]{ahn2023modification}
Ahn,~W.; Triana,~J.~F.; Recabal,~F.; Herrera,~F.; Simpkins,~B.~S. Modification of ground-state chemical reactivity via light--matter coherence in infrared cavities. \emph{Science} \textbf{2023}, \emph{380}, 1165--1168\relax
\mciteBstWouldAddEndPuncttrue
\mciteSetBstMidEndSepPunct{\mcitedefaultmidpunct}
{\mcitedefaultendpunct}{\mcitedefaultseppunct}\relax
\EndOfBibitem
\bibitem[Lather \latin{et~al.}(2022)Lather, Thabassum, Singh, and George]{lather2022cavity}
Lather,~J.; Thabassum,~A.~N.; Singh,~J.; George,~J. Cavity catalysis: modifying linear free-energy relationship under cooperative vibrational strong coupling. \emph{Chem. Sci.} \textbf{2022}, \emph{13}, 195--202\relax
\mciteBstWouldAddEndPuncttrue
\mciteSetBstMidEndSepPunct{\mcitedefaultmidpunct}
{\mcitedefaultendpunct}{\mcitedefaultseppunct}\relax
\EndOfBibitem
\bibitem[Mandal \latin{et~al.}(2023)Mandal, Taylor, Weight, Koessler, Li, and Huo]{mandal2023theoretical}
Mandal,~A.; Taylor,~M.~A.; Weight,~B.~M.; Koessler,~E.~R.; Li,~X.; Huo,~P. Theoretical advances in polariton chemistry and molecular cavity quantum electrodynamics. \emph{Chem. Rev.} \textbf{2023}, \emph{123}, 9786--9879\relax
\mciteBstWouldAddEndPuncttrue
\mciteSetBstMidEndSepPunct{\mcitedefaultmidpunct}
{\mcitedefaultendpunct}{\mcitedefaultseppunct}\relax
\EndOfBibitem
\bibitem[Thomas \latin{et~al.}(2020)Thomas, Jayachandran, Lethuillier-Karl, Vergauwe, Nagarajan, Devaux, Genet, Moran, and Ebbesen]{ThomasNp2020}
Thomas,~A.; Jayachandran,~A.; Lethuillier-Karl,~L.; Vergauwe,~R.~M.; Nagarajan,~K.; Devaux,~E.; Genet,~C.; Moran,~J.; Ebbesen,~T.~W. Ground state chemistry under vibrational strong coupling: dependence of thermodynamic parameters on the Rabi splitting energy. \emph{Nanophotonics} \textbf{2020}, \emph{9}, 249--255\relax
\mciteBstWouldAddEndPuncttrue
\mciteSetBstMidEndSepPunct{\mcitedefaultmidpunct}
{\mcitedefaultendpunct}{\mcitedefaultseppunct}\relax
\EndOfBibitem
\bibitem[Fowler-Wright \latin{et~al.}(2022)Fowler-Wright, Lovett, and Keeling]{fowler2022efficient}
Fowler-Wright,~P.; Lovett,~B.~W.; Keeling,~J. Efficient many-body non-Markovian dynamics of organic polaritons. \emph{Phys. Rev. Lett.} \textbf{2022}, \emph{129}, 173001\relax
\mciteBstWouldAddEndPuncttrue
\mciteSetBstMidEndSepPunct{\mcitedefaultmidpunct}
{\mcitedefaultendpunct}{\mcitedefaultseppunct}\relax
\EndOfBibitem
\bibitem[Fowler-Wright(2024)]{fowler2024mean}
Fowler-Wright,~P. Mean-field and cumulant approaches to modelling organic polariton physics. \emph{arXiv preprint arXiv:2405.09812} \textbf{2024}, \relax
\mciteBstWouldAddEndPunctfalse
\mciteSetBstMidEndSepPunct{\mcitedefaultmidpunct}
{}{\mcitedefaultseppunct}\relax
\EndOfBibitem
\bibitem[Lindoy \latin{et~al.}(2024)Lindoy, Mandal, and Reichman]{lindoy2024investigating}
Lindoy,~L.~P.; Mandal,~A.; Reichman,~D.~R. Investigating the collective nature of cavity-modified chemical kinetics under vibrational strong coupling. \emph{Nanophotonics} \textbf{2024}, \emph{13}, 2617--2633\relax
\mciteBstWouldAddEndPuncttrue
\mciteSetBstMidEndSepPunct{\mcitedefaultmidpunct}
{\mcitedefaultendpunct}{\mcitedefaultseppunct}\relax
\EndOfBibitem
\bibitem[Crespo-Otero and Barbatti(2018)Crespo-Otero, and Barbatti]{crespo2018recent}
Crespo-Otero,~R.; Barbatti,~M. Recent advances and perspectives on nonadiabatic mixed quantum--classical dynamics. \emph{Chem. Rev.} \textbf{2018}, \emph{118}, 7026--7068\relax
\mciteBstWouldAddEndPuncttrue
\mciteSetBstMidEndSepPunct{\mcitedefaultmidpunct}
{\mcitedefaultendpunct}{\mcitedefaultseppunct}\relax
\EndOfBibitem
\bibitem[Hu \latin{et~al.}(2023)Hu, Ying, and Huo]{hu2023resonance}
Hu,~D.; Ying,~W.; Huo,~P. Resonance enhancement of vibrational polariton chemistry obtained from the mixed quantum-classical dynamics simulations. \emph{J. Phys. Chem. Lett.} \textbf{2023}, \emph{14}, 11208--11216\relax
\mciteBstWouldAddEndPuncttrue
\mciteSetBstMidEndSepPunct{\mcitedefaultmidpunct}
{\mcitedefaultendpunct}{\mcitedefaultseppunct}\relax
\EndOfBibitem
\bibitem[Lindoy \latin{et~al.}(2023)Lindoy, Mandal, and Reichman]{lindoy2023quantum}
Lindoy,~L.~P.; Mandal,~A.; Reichman,~D.~R. Quantum dynamical effects of vibrational strong coupling in chemical reactivity. \emph{Nat. Comm.} \textbf{2023}, \emph{14}, 2733\relax
\mciteBstWouldAddEndPuncttrue
\mciteSetBstMidEndSepPunct{\mcitedefaultmidpunct}
{\mcitedefaultendpunct}{\mcitedefaultseppunct}\relax
\EndOfBibitem
\bibitem[Mannouch and Richardson(2023)Mannouch, and Richardson]{mannouch2023mapping}
Mannouch,~J.~R.; Richardson,~J.~O. A mapping approach to surface hopping. \emph{J. Chem. Phys.} \textbf{2023}, \emph{158}\relax
\mciteBstWouldAddEndPuncttrue
\mciteSetBstMidEndSepPunct{\mcitedefaultmidpunct}
{\mcitedefaultendpunct}{\mcitedefaultseppunct}\relax
\EndOfBibitem
\bibitem[Runeson and Manolopoulos(2023)Runeson, and Manolopoulos]{runeson2023multi}
Runeson,~J.~E.; Manolopoulos,~D.~E. A multi-state mapping approach to surface hopping. \emph{J. Chem. Phys.} \textbf{2023}, \emph{159}\relax
\mciteBstWouldAddEndPuncttrue
\mciteSetBstMidEndSepPunct{\mcitedefaultmidpunct}
{\mcitedefaultendpunct}{\mcitedefaultseppunct}\relax
\EndOfBibitem
\bibitem[Tanimura(1990)]{tanimura1990nonperturbative}
Tanimura,~Y. Nonperturbative expansion method for a quantum system coupled to a harmonic-oscillator bath. \emph{Phys. Rev. A} \textbf{1990}, \emph{41}, 6676\relax
\mciteBstWouldAddEndPuncttrue
\mciteSetBstMidEndSepPunct{\mcitedefaultmidpunct}
{\mcitedefaultendpunct}{\mcitedefaultseppunct}\relax
\EndOfBibitem
\bibitem[Tanimura(2006)]{tanimura2006stochastic}
Tanimura,~Y. Stochastic Liouville, Langevin, Fokker--Planck, and master equation approaches to quantum dissipative systems. \emph{J. Phys. Soc. Jap.} \textbf{2006}, \emph{75}, 082001\relax
\mciteBstWouldAddEndPuncttrue
\mciteSetBstMidEndSepPunct{\mcitedefaultmidpunct}
{\mcitedefaultendpunct}{\mcitedefaultseppunct}\relax
\EndOfBibitem
\bibitem[Xu \latin{et~al.}(2005)Xu, Cui, Li, Mo, and Yan]{xu2005exact}
Xu,~R.-X.; Cui,~P.; Li,~X.-Q.; Mo,~Y.; Yan,~Y. Exact quantum master equation via the calculus on path integrals. \emph{J. Chem. Phys.} \textbf{2005}, \emph{122}\relax
\mciteBstWouldAddEndPuncttrue
\mciteSetBstMidEndSepPunct{\mcitedefaultmidpunct}
{\mcitedefaultendpunct}{\mcitedefaultseppunct}\relax
\EndOfBibitem
\bibitem[Xu and Yan(2007)Xu, and Yan]{xu2007dynamics}
Xu,~R.-X.; Yan,~Y. Dynamics of quantum dissipation systems interacting with bosonic canonical bath: Hierarchical equations of motion approach. \emph{Phys. Rev. E} \textbf{2007}, \emph{75}, 031107\relax
\mciteBstWouldAddEndPuncttrue
\mciteSetBstMidEndSepPunct{\mcitedefaultmidpunct}
{\mcitedefaultendpunct}{\mcitedefaultseppunct}\relax
\EndOfBibitem
\bibitem[Lawrence \latin{et~al.}(2024)Lawrence, Mannouch, and Richardson]{lawrence2024size}
Lawrence,~J.~E.; Mannouch,~J.~R.; Richardson,~J.~O. A size-consistent multi-state mapping approach to surface hopping. \emph{J. Chem. Phys.} \textbf{2024}, \emph{160}, 244112\relax
\mciteBstWouldAddEndPuncttrue
\mciteSetBstMidEndSepPunct{\mcitedefaultmidpunct}
{\mcitedefaultendpunct}{\mcitedefaultseppunct}\relax
\EndOfBibitem
\bibitem[Cotton and Miller(2019)Cotton, and Miller]{cotton2019trajectory}
Cotton,~S.~J.; Miller,~W.~H. Trajectory-adjusted electronic zero point energy in classical Meyer-Miller vibronic dynamics: Symmetrical quasiclassical application to photodissociation. \emph{J. Chem. Phys.} \textbf{2019}, \emph{150}\relax
\mciteBstWouldAddEndPuncttrue
\mciteSetBstMidEndSepPunct{\mcitedefaultmidpunct}
{\mcitedefaultendpunct}{\mcitedefaultseppunct}\relax
\EndOfBibitem
\bibitem[Meyera and Miller(1979)Meyera, and Miller]{meyera1979classical}
Meyera,~H.-D.; Miller,~W.~H. A classical analog for electronic degrees of freedom in nonadiabatic collision processes. \emph{J. Chem. Phys.} \textbf{1979}, \emph{70}, 3214--3223\relax
\mciteBstWouldAddEndPuncttrue
\mciteSetBstMidEndSepPunct{\mcitedefaultmidpunct}
{\mcitedefaultendpunct}{\mcitedefaultseppunct}\relax
\EndOfBibitem
\bibitem[Stock and Thoss(1997)Stock, and Thoss]{stock1997semiclassical}
Stock,~G.; Thoss,~M. Semiclassical description of nonadiabatic quantum dynamics. \emph{Phys. Rev. Lett.} \textbf{1997}, \emph{78}, 578\relax
\mciteBstWouldAddEndPuncttrue
\mciteSetBstMidEndSepPunct{\mcitedefaultmidpunct}
{\mcitedefaultendpunct}{\mcitedefaultseppunct}\relax
\EndOfBibitem
\bibitem[Thoss and Stock(1999)Thoss, and Stock]{thoss1999mapping}
Thoss,~M.; Stock,~G. Mapping approach to the semiclassical description of nonadiabatic quantum dynamics. \emph{Phys. Rev. A} \textbf{1999}, \emph{59}, 64\relax
\mciteBstWouldAddEndPuncttrue
\mciteSetBstMidEndSepPunct{\mcitedefaultmidpunct}
{\mcitedefaultendpunct}{\mcitedefaultseppunct}\relax
\EndOfBibitem
\bibitem[Miller and Cotton(2016)Miller, and Cotton]{miller2016classical}
Miller,~W.~H.; Cotton,~S.~J. Classical molecular dynamics simulation of electronically non-adiabatic processes. \emph{Faraday Discuss.} \textbf{2016}, \emph{195}, 9--30\relax
\mciteBstWouldAddEndPuncttrue
\mciteSetBstMidEndSepPunct{\mcitedefaultmidpunct}
{\mcitedefaultendpunct}{\mcitedefaultseppunct}\relax
\EndOfBibitem
\bibitem[Runeson and Richardson(2019)Runeson, and Richardson]{runeson2019spin}
Runeson,~J.~E.; Richardson,~J.~O. Spin-mapping approach for nonadiabatic molecular dynamics. \emph{J. Chem. Phys.} \textbf{2019}, \emph{151}\relax
\mciteBstWouldAddEndPuncttrue
\mciteSetBstMidEndSepPunct{\mcitedefaultmidpunct}
{\mcitedefaultendpunct}{\mcitedefaultseppunct}\relax
\EndOfBibitem
\bibitem[Runeson and Richardson(2020)Runeson, and Richardson]{runeson2020generalized}
Runeson,~J.~E.; Richardson,~J.~O. Generalized spin mapping for quantum-classical dynamics. \emph{J. Chem. Phys.} \textbf{2020}, \emph{152}\relax
\mciteBstWouldAddEndPuncttrue
\mciteSetBstMidEndSepPunct{\mcitedefaultmidpunct}
{\mcitedefaultendpunct}{\mcitedefaultseppunct}\relax
\EndOfBibitem
\bibitem[Tully(1990)]{tully1990molecular}
Tully,~J.~C. Molecular dynamics with electronic transitions. \emph{J. Chem. Phys.} \textbf{1990}, \emph{93}, 1061--1071\relax
\mciteBstWouldAddEndPuncttrue
\mciteSetBstMidEndSepPunct{\mcitedefaultmidpunct}
{\mcitedefaultendpunct}{\mcitedefaultseppunct}\relax
\EndOfBibitem
\bibitem[Wang \latin{et~al.}(2014)Wang, Trivedi, and Prezhdo]{wang2014global}
Wang,~L.; Trivedi,~D.; Prezhdo,~O.~V. Global flux surface hopping approach for mixed quantum-classical dynamics. \emph{J. Chem. Theory Comput.} \textbf{2014}, \emph{10}, 3598--3605\relax
\mciteBstWouldAddEndPuncttrue
\mciteSetBstMidEndSepPunct{\mcitedefaultmidpunct}
{\mcitedefaultendpunct}{\mcitedefaultseppunct}\relax
\EndOfBibitem
\bibitem[Lawrence \latin{et~al.}(2024)Lawrence, Mannouch, and Richardson]{lawrence2024recovering}
Lawrence,~J.~E.; Mannouch,~J.~R.; Richardson,~J.~O. Recovering Marcus theory rates and beyond without the need for decoherence corrections: The mapping approach to surface hopping. \emph{J. Phys. Chem. Lett.} \textbf{2024}, \emph{15}, 707--716\relax
\mciteBstWouldAddEndPuncttrue
\mciteSetBstMidEndSepPunct{\mcitedefaultmidpunct}
{\mcitedefaultendpunct}{\mcitedefaultseppunct}\relax
\EndOfBibitem
\bibitem[Mandal \latin{et~al.}(2020)Mandal, Montillo~Vega, and Huo]{mandal2020polarized}
Mandal,~A.; Montillo~Vega,~S.; Huo,~P. Polarized Fock states and the dynamical Casimir effect in molecular cavity quantum electrodynamics. \emph{J. Phys. Chem. Lett.} \textbf{2020}, \emph{11}, 9215--9223\relax
\mciteBstWouldAddEndPuncttrue
\mciteSetBstMidEndSepPunct{\mcitedefaultmidpunct}
{\mcitedefaultendpunct}{\mcitedefaultseppunct}\relax
\EndOfBibitem
\bibitem[Chandler(1978)]{chandler1978statistical}
Chandler,~D. Statistical mechanics of isomerization dynamics in liquids and the transition state approximation. \emph{J. Chem. Phys.} \textbf{1978}, \emph{68}, 2959--2970\relax
\mciteBstWouldAddEndPuncttrue
\mciteSetBstMidEndSepPunct{\mcitedefaultmidpunct}
{\mcitedefaultendpunct}{\mcitedefaultseppunct}\relax
\EndOfBibitem
\bibitem[Chandler(1987)]{chandler1987introduction}
Chandler,~D. Introduction to modern statistical. \emph{Mechanics. Oxford University Press, Oxford, UK} \textbf{1987}, \emph{5}, 11\relax
\mciteBstWouldAddEndPuncttrue
\mciteSetBstMidEndSepPunct{\mcitedefaultmidpunct}
{\mcitedefaultendpunct}{\mcitedefaultseppunct}\relax
\EndOfBibitem
\bibitem[Xie \latin{et~al.}(2013)Xie, Bai, Zhu, and Shi]{xie2013calculation}
Xie,~W.; Bai,~S.; Zhu,~L.; Shi,~Q. Calculation of electron transfer rates using mixed quantum classical approaches: Nonadiabatic limit and beyond. \emph{J. Phys. Chem. A} \textbf{2013}, \emph{117}, 6196--6204\relax
\mciteBstWouldAddEndPuncttrue
\mciteSetBstMidEndSepPunct{\mcitedefaultmidpunct}
{\mcitedefaultendpunct}{\mcitedefaultseppunct}\relax
\EndOfBibitem
\bibitem[Ying and Huo(2023)Ying, and Huo]{ying2023resonance}
Ying,~W.; Huo,~P. Resonance theory and quantum dynamics simulations of vibrational polariton chemistry. \emph{J. Chem. Phys.} \textbf{2023}, \emph{159}\relax
\mciteBstWouldAddEndPuncttrue
\mciteSetBstMidEndSepPunct{\mcitedefaultmidpunct}
{\mcitedefaultendpunct}{\mcitedefaultseppunct}\relax
\EndOfBibitem
\bibitem[Walters \latin{et~al.}(2017)Walters, Allen, and Makri]{walters2017direct}
Walters,~P.~L.; Allen,~T.~C.; Makri,~N. Direct determination of discrete harmonic bath parameters from molecular dynamics simulations. \emph{J. Comp. Chem.} \textbf{2017}, \emph{38}, 110--115\relax
\mciteBstWouldAddEndPuncttrue
\mciteSetBstMidEndSepPunct{\mcitedefaultmidpunct}
{\mcitedefaultendpunct}{\mcitedefaultseppunct}\relax
\EndOfBibitem
\bibitem[Runeson \latin{et~al.}(2024)Runeson, Fay, and Manolopoulos]{runeson2024exciton}
Runeson,~J.~E.; Fay,~T.~P.; Manolopoulos,~D.~E. Exciton dynamics from the mapping approach to surface hopping: comparison with F{\"o}rster and Redfield theories. \emph{Phys. Chem. Chem. Phys.} \textbf{2024}, \emph{26}, 4929--4938\relax
\mciteBstWouldAddEndPuncttrue
\mciteSetBstMidEndSepPunct{\mcitedefaultmidpunct}
{\mcitedefaultendpunct}{\mcitedefaultseppunct}\relax
\EndOfBibitem
\bibitem[Granucci \latin{et~al.}(2001)Granucci, Persico, and Toniolo]{granucci2001direct}
Granucci,~G.; Persico,~M.; Toniolo,~A. Direct semiclassical simulation of photochemical processes with semiempirical wave functions. \emph{J. Chem. Phys.} \textbf{2001}, \emph{114}, 10608--10615\relax
\mciteBstWouldAddEndPuncttrue
\mciteSetBstMidEndSepPunct{\mcitedefaultmidpunct}
{\mcitedefaultendpunct}{\mcitedefaultseppunct}\relax
\EndOfBibitem
\end{mcitethebibliography}


\end{document}